\documentclass[aps,prb,superscriptaddress,twocolumn,longbibliography]{revtex4-1}
\usepackage{bbm}
\usepackage{graphicx}
\usepackage{dcolumn}
\usepackage{bm}
\usepackage{subfigure}
\usepackage{amsmath}
\usepackage{feynmf}
\usepackage{hyperref}
\usepackage{color}
\usepackage{braket}
\usepackage{amssymb}

\usepackage{attachfile}

\newcommand{\bk}{\boldsymbol k}

\usepackage{times}
\newcommand{\Exp}{\text{Exp}}
\newcommand{\Tr}{\text{Tr}}

\begin{document}
\title{Multifold Majorana corner modes  arising from multiple pairs of helical edge states}

\author{Zhiwei Yin}
\thanks{These authors contributed equally to this work.}
\affiliation{Department of Modern Physics, University of Science and Technology of China, Hefei, 230026, China}

\author{Haoshu Li}
\thanks{These authors contributed equally to this work.}
\affiliation{Department of Physics, University of Science and Technology of China, Hefei, 230026, China}

\author{Zhongbo Yan}
\email{Contact author: yanzhb5@mail.sysu.edu.cn}
\affiliation{Guangdong Provincial Key Laboratory of Magnetoelectric Physics and Devices, State Key Laboratory of Optoelectronic Materials and Technologies,
School of Physics, Sun Yat-sen University, Guangzhou 510275, China}

\author{Shaolong Wan}
\affiliation{Department of Modern Physics, University of Science and Technology of China, Hefei, 230026, China}

\date{\today}

\begin{abstract}
Quantum spin Hall insulators with a pair of helical edge states and proximity-induced superconductivity have
been shown to support second-order topological superconductors with Majorana corner modes. As the Majorana corner
modes are originated from the helical edge states of the quantum spin Hall insulators, whether
quantum spin Hall insulators with multiple pairs of helical edge states and proximity-induced superconductivity
can give rise to second-order topological superconductors with multifold Majorana corner modes
is an interesting question to address. In this work,
we consider a quantum spin Hall insulator with two pairs of helical edge states. We
find robust twofold Majorana corner modes can be achieved when
the helical edge states are gapped by a combined action of a magnetic exchange field
and an $s$-wave pairing, or an $s+p$ mixed-parity pairing.
The stability of two Majorana zero modes per corner under the action of magnetic exchange fields
is attributed to the protection from the chiral symmetry. Our study reveals that heterostructures composed of superconductors
and quantum spin Hall insulators with multiple pairs of helical edge states could serve as a platform to pursue
multifold Majorana corner modes.
\end{abstract}

\maketitle

\section{Introduction}

Majorana fermions are famous in particle physics for being their own antiparticles.
Their zero-dimensional (0D) counterpart in condensed matter systems,
known as Majorana zero mode (MZM),
has attracted great interest over the past two decades due to
its exotic properties and prospect in implementing topological quantum computation\cite{nayak2008review,leijnse2012introduction,alicea2012new,Beenakker2013,stanescu2013majorana,
Elliott2015,sarma2015majorana,sato2016majorana,Aguado2017,Flensberg2021review,Jack2021review,Hsu2021review}. MZMs
were first unveiled to emerge at the ends of 1D $p$-wave superconductors \cite{kitaev2001unpaired}
and the vortex cores of 2D chiral $p$-wave superconductors\cite{Volovik1999MZM,read2000}. Later, with the
advent of topological insulators, a diversity of more accessible protocols were discovered\cite{fu2008,Fu2009junction,lutchyn2010,oreg2010helical,sau2010,alicea2010},
accompanying with the succeeding observations of signatures pointing to MZMs in
numerous material platforms, ranging from 1D superconducting nanowires to 3D iron-based
superconductors with an inverted band structure\cite{Mourik2012MZM,Nadj2014MZM,Sun2016Majorana,wang2018evidence,kong2019half,Fornieri2019,Ren2019Planar}.

In dimensions $d\geq2$, the realization of MZMs commonly relies on
the cooperation of momentum-space topology and real-space topological
defects, such as vortices\cite{Hosur2011MZM,Yan2017MZM,Yan2020vortex,Giwa2021MZM,Hu2023vortex}, dislocations\cite{Hughes2014dislocation,Das2023AI,Hu2024dislocation,Zhu2024dislocation} and disclinations\cite{Teo2013disclination,Benalcazar2014}. In the
past few years, a remarkable finding is that MZMs can also directly appear at the boundary of a
higher-dimensional superconducting system with higher-order topology. To be specific,
an $n$th-order topological superconductor in $d$ dimensions ($n\leq d$)
refers to a phase hosting gapless boundary states of codimension equal to $n$\cite{Benalcazar2017,Benalcazar2017prb,Schindler2018,Song2017,Langbehn2017}. When $n=d$, MZMs are shown
to appear at the corners of the $d$D system.
As topological defects are no more needed,
higher-order topological superconductors (HOTSCs) referring to $n\geq2$ are rather appealing for the realization
and exploitation of MZMs\cite{Zhang2020SOTSCa,Zhang2020tunable,Pahomi2020braiding,Lapa2021}. Theoretical studies have found that there are many ways to realize HOTSCs.
Representative protocols for the realizaton of MZMs in the context of HOTSCs include heterostructures composed of topological insulators and superconductors\cite{Yan2018hosc,Wang2018hosc,Liu2018hosc,Hsu2018hosc,Pan2019SOTSC,Wu2020SOTSC,Yan2019hoscb,Zhu2022sublattice,Chen2023MCM},
heterostructures composed of obstructed atomic insulator and superconductors\cite{Li2021BTSC,Sheng2024OAI}, $\pi$-junction Rashba layers\cite{Volpez2019SOTSC,Laubscher2020}, superconductors with topological band structures and unconventional even-parity  pairings\cite{Zhang2019hoscb,wu2020boundaryobstructedb,Chen2021SOTSC,Qin2022hosc,Zhang2024hosc},
superconductors with mixed-parity pairings\cite{Wangyuxuan2018hosc,Wuzhigang2019hosc,Ikegaya2021hosc,Manna2022fractal}, spin-orbit coupled superconductors with $s+id$ pairings\cite{Zhu2019mixed,Majid2020hoscb}, odd-parity superconductors\cite{Zhu2018hosc,Yan2019hosca,Hsu2020hosc,Ahn2020hosc,Zhu2023AM} and twisted bilayer systems\cite{Li2023twisted,Chew2023hotsc}.
All these theoretical protocols, however, are still waiting for an experimental realization.

Up to now, theoretical studies on the realization of MZMs in HOTSCs without time-reversal symmetry
mainly focus on isolated MZMs. The underlying reason is that MZMs normally follow a $Z_{2}$ classification.
Namely, although an isolated MZM is robust due to the particle-hole symmetry (PHS), two overlapped MZMs will hybrid and lead to the losing
of their self-conjugate nature if there is no additional symmetry prohibiting them from coupling.
Recently, several works have uncovered
that a 2D second-order topological insulator with chiral symmetry
can support multiple zero-energy bound states per corner\cite{Benalcazar2022CSHOTI,Wang2023MZCM,Li2023CSHOTI,Li2023CSHOTIb,Li2024CSHOTI,Yang2024CSHOTI}. In principle,
this scenario can also be generalized to superconductors
as the prohibition of hybridization by chiral symmetry does not depend on
the nature of the zero-energy states\cite{Mondal2024,Zhu2024CSHOTSC}. However,
how to realize a HOTSC with multiple MZMs per corner [termed as multifold Majorana corner modes (MCMs)]
in a realistic way remains unexplored.

Before introducing our protocol,  a remarkable difference between
insulators and superconductors is worthy of emphasizing. That is,
the chiral symmetry in insulators can be an independent symmetry,
whereas in superconductors it is always connected to the time-reversal symmetry (TRS). This is because the Bogoliubov-de Gennes (BdG) Hamiltonians
describing superconductors at the mean-field level always have PHS, and the product
of TRS and PHS gives the chiral symmetry.
Since there exist two types of TRS, namely, TRS for integer spin (including effectively spinless cases)
and half-integer spin\cite{schnyder2008}, there also exist two types of chiral symmetry. In this work, we will show
that only the chiral symmetry tied to the first-type TRS can protect as many  MZMs as one wants.

In the protocol of realizing 2D second-order topological superconductors (SOTSCs) based on heterostructures
composed of topological insulators and superconductors, an isolated MZM per corner is found to appear
if a pair of helical edge states is gapped by the combined action of superconductivity and magnetic exchange field
(or Zeeman field)\cite{Liu2018hosc,Pan2019SOTSC,Wu2020SOTSC}, or a Majorana Kramers pair per corner is found to appear
if the pair of helical edge states is gapped by unconventional superconductivity without the breaking
of TRS\cite{Yan2018hosc,Wang2018hosc}. As the MCMs arise from the helical edge states,
it is sensible to conjecture that multifold MCMs may be achieved
if there exist multiple pairs of helical edge states and they are gapped by superconductivity
and magnetic exchange field in an appropriate way. In this work, we consider a quantum spin Hall insulator with two pairs of helical edge states
to demonstrate the conjecture. To be specific, we consider two ways to gap out the helical edge states,
The first way is a combined action of a magnetic exchange field and an $s$-wave pairing, and the second
way is an action of an $s+p$ mixed-parity pairing. We find that both approaches
can realize SOTCSs supporting twofold MCMs even when the spinful
TRS enforcing Kramers degeneracy is broken, as long as the BdG Hamiltonians
have an effective spinless TRS.

The rest of the paper is organized as follows. In Sec.\ref{II}, we discuss
a quantum spin Hall insulator with two pairs of helical edge states.
In Sec.\ref{III}, we investigate the influence of a magnetic exchange field and an $s$-wave pairing
on the two pairs of helical edge states. We develop an edge-state theory to show
how twofold MCMs arise. The multipole chiral number, a
bulk topological invariant, is also calculated to see the match of the prediction based on
the edge-state theory and bulk topological invariant. In Sec.\ref{IV},
we explore the influence of an $s+p$ mixed-parity pairing
on the helical edge states. The rise and the stability of twofold MCMs in
this case are also elucidated in terms of similar edge-state theory and
multipole chiral number. In Sec.\ref{V},  we discuss generalizations and potential experimental realizations,
and then conclude the paper. Some calculation details are relegated
to the Appendixes.

\section{Theoretical model of the quantum spin Hall insulator}\label{II}
We start with the following Hamiltonian $H= \sum_{\bm{k}} \psi_{\boldsymbol{k}}^{\dagger} H_{N}(\boldsymbol{k}) \psi_{\boldsymbol{k}}$,
with $\psi_{\boldsymbol{k}}=\left(c_{\boldsymbol{k}, a, \uparrow}, c_{\boldsymbol{k}, b, \uparrow}, c_{\boldsymbol{k}, a, \downarrow}, c_{\boldsymbol{k}, b, \downarrow}\right)^T\,$ and\cite{PhysRevB.94.235111}
\begin{equation}
    \begin{aligned}
        H_{N}(\bk) &= (m-2 t_x\cos k_x - 2 t_y\cos k_y)\sigma_z  \\
                &+2\lambda \left(\cos k_x -\cos k_y \right)\sigma_x\\
                &+2\lambda \sin k_x\sin k_y\sigma_y s_y\,,
    \end{aligned}
    \label{eq:baseTI}
    \end{equation}
where $s_i$ and $\sigma_i$ are the Pauli matrices acting on the spin and orbital degrees of freedom, respectively; $t_x$, $t_y$ and $\lambda$ are hopping amplitudes, and $m$ determines the band inversion. Throughout this paper, the lattice constant is set to unity and all identity matrices are made implicit for notational simplicity, and without loss
of generality, the parameters $m$, $t_x$, $t_y$ and $\lambda$ are assumed to be non-negative for the convenience of discussion.

\begin{figure}[t!]
    \centering
    \includegraphics[width=0.96\columnwidth]{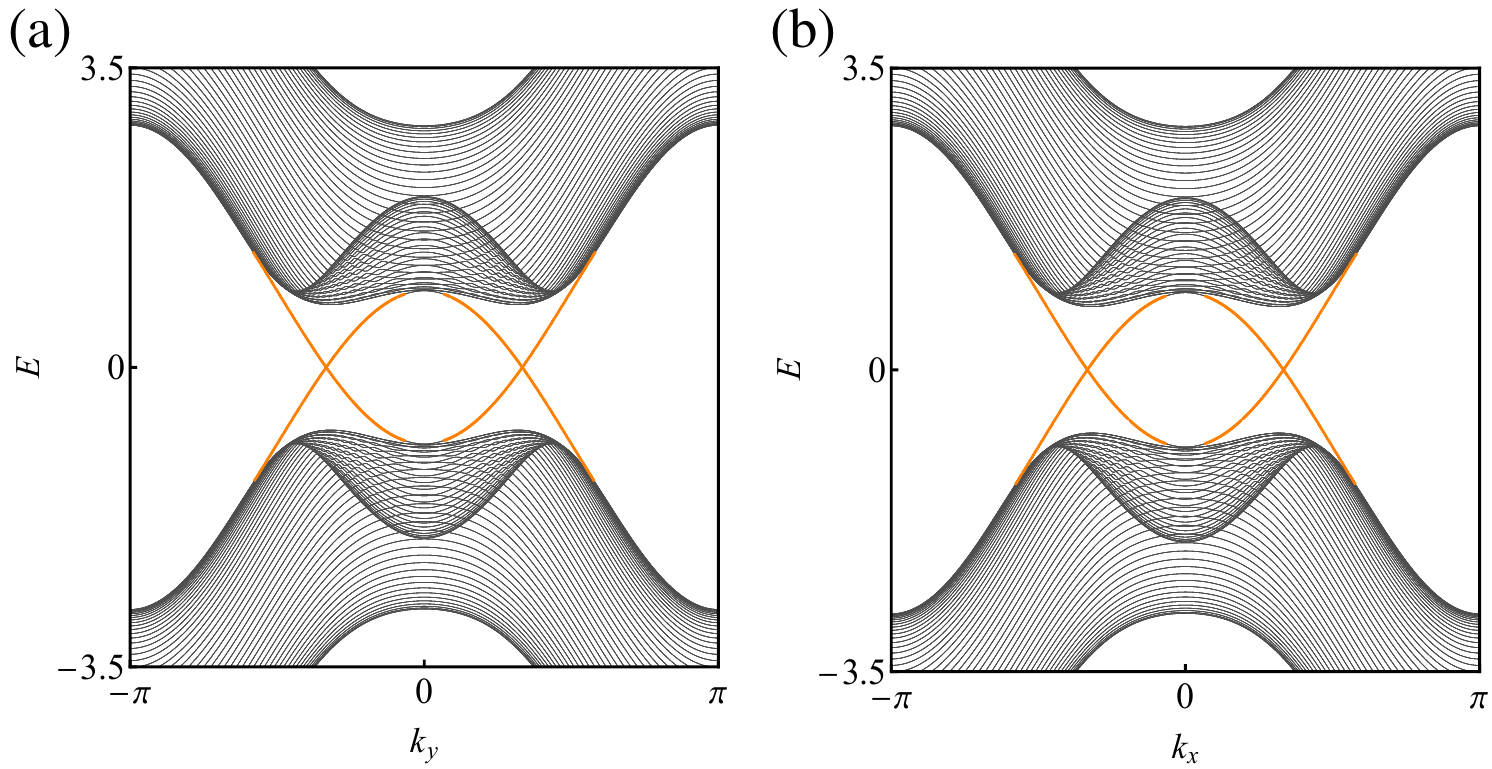}
    \caption{Energy spectra of the normal state for a sample of  cylindrical  geometry. Two pairs of gapless helical edge states are present on both types of boundaries, traversing the bulk gap, as indicated by the solid orange lines. (a)[(b)] Energy spectra with open boundary conditions in the $x$($y$)-direction, showing edge states localized at the left and right (top and bottom) boundaries.  Common parameters are $m=2$, $t_x=t_y=1$ and $\lambda=0.5$. The number of lattice sites along the direction with open boundary conditions is $N = 50$.}
    \label{fig.1}
\end{figure}

It is evident that the Hamiltonian possesses TRS, $\mathcal T H_{N}(\bk) \mathcal T^{-1} = H_{N}(-\bk)$, where the time-reversal operator is given by  $\mathcal{T}=is_y \mathcal{K}\,$, with $\mathcal{K}$ the complex conjugate operator. In this paper, for the sake of discussion convenience, we will refer TRS with  $\mathcal{T}^{2}=-1$ (1) as spinful (spinless) TRS.
The Hamiltonian also has inversion symmetry. Since the involved orbitals of the Hamiltonian have the same parity, the inversion operator is the identity matrix, i.e., $P=\textbf{I}$. This form
indicates that the insulator phase will always be $Z_{2}$ trivial according to the Fu-Kane criterion which uses the parity eigenvalues of occupied states at the time-reversal invariant momenta\cite{fu2007a}. However, due to the existence of a $U(1)$ spin rotation symmetry which renders the spin component $s_y$ conserved, the Hamiltonian is extended to follow a
$Z$ classification and characterized by the so-called spin Chern number\cite{PhysRevLett.97.036808}.
To be specific, according to the two eigenvalues of $s_{y}$ the Hamiltonian can be decomposed into two decoupled sectors, i.e., $H_{N}(\bk)=H_{+1}(\bk)\oplus H_{-1}(\bk)$, where
\begin{equation}
    \begin{aligned}
        H_{s=\pm1}(\bk) &=(m-2 t_x\cos k_x - 2 t_y\cos k_y) \sigma_z \\
                &\phantom{+}\;+2\lambda \left(\cos k_x -\cos k_y \right)\sigma_x  \\
                &\phantom{+}\;+2s\lambda \sin k_x\sin k_y\sigma_y.
    \end{aligned}
\end{equation}
Expressing the Hamiltonian as $H_{s}(\bk)=\bm{d}_{s}(\bk)\cdot\boldsymbol{\sigma}$
with $\bm{d}_{s}(\bk) =(2\lambda (\cos k_x -\cos k_y),\,2s\lambda \sin k_x\sin k_y,\,m-2 t_x\cos k_x - 2 t_y\cos k_y)$, the Chern number characterizing the two-band Hamiltonian for each spin sector is simply given by\cite{qi2005}
\begin{equation}
    \begin{aligned}
    C_{s} &=\frac{1}{4 \pi} \int_{B Z} \frac{\boldsymbol{d}_{s}(\boldsymbol{k})         \cdot\left(\partial_{k_x} \boldsymbol{d}_{s}(\boldsymbol{k}) \times \partial_{k_y} \boldsymbol{d}_{s}(\boldsymbol{k})\right)}{|\bm d(\boldsymbol{k})|^{3}} d^2 k\\
     &= \left\{\begin{array}{cl}
        2s,& |m|<2(t_x+t_y)\,, \\
        0, & |m|>2(t_x+t_y)\,.
        \end{array}\right.
    \end{aligned}
    \end{equation}
In the case of that $|m|<2(t_x+t_y)$, the Chern numbers for the two spin sectors are
equal to $\pm2$, yielding the spin
Chern number $C_{\rm spin}= (C_{+1}-C_{-1})/2=2$.  which corresponds to the realization of a quantum spin Hall insulator with two pairs of helical edge states.
In Fig.\ref{fig.1}, we show
the energy spectra of the Hamiltonian for a sample of
cylindrical geometry.
Apparently, there are two pairs of gapless helical edge states along both the $x$-normal  and $y$-normal edges, agreeing with the expectation from the bulk-boundary correspondence.

\section{Second-order topological superconductors with twofold Majorana corner modes based on even-parity superconductivity}\label{III}
\subsection{Bulk Hamiltonian}\label{sec: Ham1}

In this section, we consider putting the  quantum spin Hall insulator in proximity
to an $s$-wave superconductor and assume
that the TRS is further broken by a magnetic exchange field.  In the Nambu basis $\Psi_{\bk}=(\psi_{\bk}, (\psi_{-\bk}^{\dagger})^{T})$, the BdG Hamiltonian describing the system is given by $\hat{H}=\frac{1}{2}\sum_{\bm{k}} \Psi_{\bm{k}}^{\dagger} H(\bm{k}) \Psi_{\bm{k}}$, where
\begin{align}\label{eq:TBSH}
    H(\boldsymbol{k})=\; & (m-2 t_x \cos k_x-2 t_y \cos k_y) \tau_z \sigma_z-\mu \tau_z   \nonumber\\
    & +2 \lambda\left(\cos k_x-\cos k_y\right) \tau_z \sigma_x  \nonumber\\
    & +2 \lambda \sin k_x \sin k_y \tau_z \sigma_y s_y \\
    & +2 \eta \sin k_x \tau_z \sigma_y s_x+\Delta_s \tau_y  s_y \,.\nonumber
\end{align}
Here the new set of Pauli matrices $\tau_i$ acts on the particle-hole degrees of freedom,  $\mu$ represents the chemical potential, $\Delta_s$ denotes the proximity-induced $s$-wave pairing amplitude, and the $\eta$ term describes a momentum-dependent exchange field
which breaks both spinful TRS and inversion symmetry but preserves their combination, which can arise from a
magnetic toroidal order\cite{Wu2024BFS}. Without loss of generality, below $\Delta_s$ and $\eta$ are also assumed to be non-negative for the sake of discussion convenience.

A common approach to generate a second-order topological phase in 2D is to gap out the gapless edge states of a first-order topological insulator and form Dirac-mass domain walls at some places of the boundary. For the quantum spin Hall insulator, the helical edge states can be gapped by
both magnetic exchange fields and superconducting pairings\cite{Fu2009junction,Fleckenstein2021}. In the Hamiltonian Eq.(\ref{eq:TBSH}), both the $\eta$ term and the $s$-wave pairing term can introduce
Dirac mass terms and open a gap to the helical edge states. Notably, the Dirac mass terms
arising from the exchange field and superconductivity are competing in nature.
Considering two intersecting edges, if the Dirac mass terms induced by the exchange field
and superconductivity dominate on different edges, their intersection, the corner, will be a Dirac-mass domain wall hosting MZMs\cite{Wu2020SOTSC}. To see whether such a scenario occurs here, we first carry out numerical calculations to determine the impact of the $\eta$ term and the $s$-wave pairing term on the helical edge states.
\begin{figure}[]
    \centering
    \includegraphics[width=0.96\columnwidth]{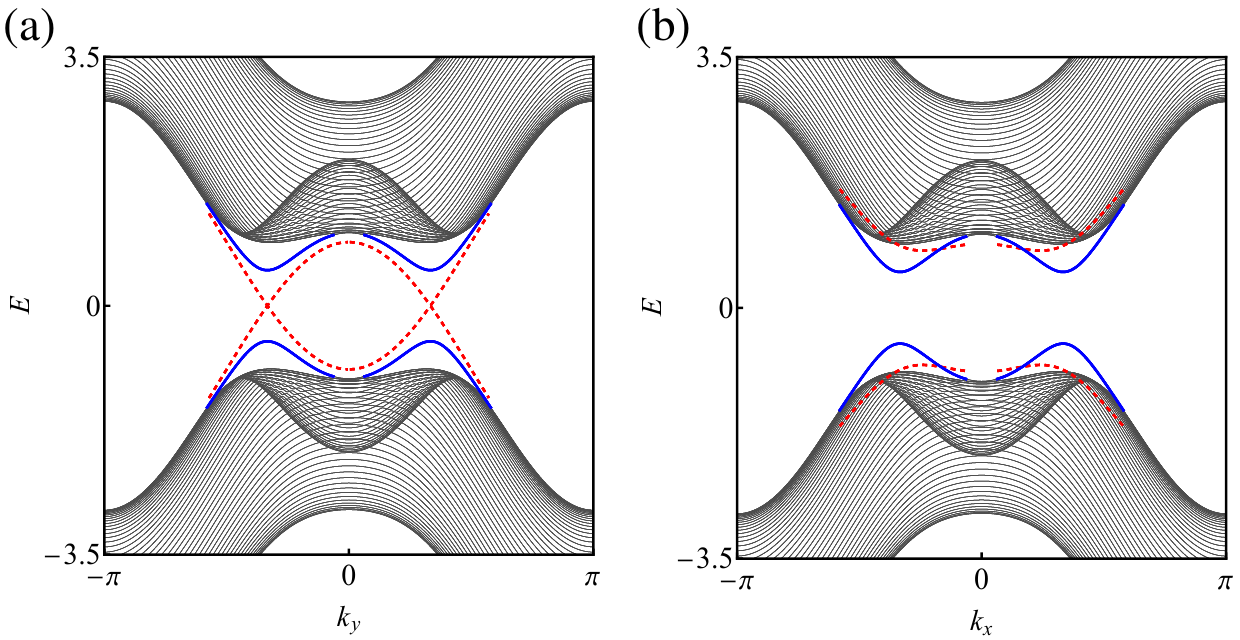}
    \caption{Energy spectra for a sample of cylindrical geometry. The red dashed lines denote the edge spectra for the case where only the $\eta$ term is present ($\Delta_s=0,\eta=0.5$). The blue solid lines correspond to the spectra when only the $s$-wave pairing term is included ($\Delta_s=0.5,\eta=0$). The bulk spectra (black curves) have little difference for these two cases, with the displayed spectra corresponding to $\Delta_s=0.5,\eta=0$. (a) The plot shows the energy spectra with open boundary conditions in the $x$ direction. The edge states are gapped by the $s$-wave pairing term, but remain gapless for the $\eta$ term. (b) The plot shows the energy spectra with open boundary conditions in the $y$ direction. Both of the two terms open the edge gap. The remaining parameters are $m=2$, $t_x=t_y=1$, $\lambda=0.5$ and $\mu=0$. }
    \label{fig.2}
\end{figure}

We focus on $\mu=0$ and first consider
$\Delta_{s}=0$. The energy spectra are shown in Fig. \ref{fig.2}. The impact of the $\eta$ term on the helical edge states is found to have a dramatic dependence on the orientation of the edges. It has little effect on the helical edge states along the $x$-normal edges but, in contrast, gaps out the two pairs of helical edge states along the $y$-normal edges, as illustrated by the red dashed lines in Figs. \ref{fig.2}(a) and  \ref{fig.2}(b). When $\eta=0$ and $\Delta_{s}\neq0$,  the energy spectra are depicted by the solid blue lines in Fig. \ref{fig.2}. Unlike the $\eta$ term, the $s$-wave pairing term opens a uniform energy gap on both $x$-normal and $y$-normal edges. This is expected since the $s$-wave pairing term is on-site and respects all symmetries of the lattice.

Based on the above results, it can be concluded that the Dirac mass on the $x$-normal edges is primarily influenced by the $s$-wave pairing term. On the $y$-normal edges, the edge gap is affected by both the $\eta$ term and the $s$-wave pairing term. Denote the two Dirac mass terms induced by
them as $M_\eta$ and $M_\Delta$, respectively. When $M_\eta$ exceeds $M_\Delta$, the Dirac mass on the y-normal edges is dominated by $M_\eta$, while the Dirac mass on the $x$-normal edges is always dominated by $M_\Delta$ as $M_\eta=0$. Therefore, according to the Jackiw-Rebbi theory\cite{jackiw1976b}, the criterion for the presence of Dirac-mass domain walls at the four corners which host MZMs is simply $|M_\eta|>|M_\Delta|$.

\begin{figure}[]
    \centering
    \includegraphics[width=0.96\columnwidth]{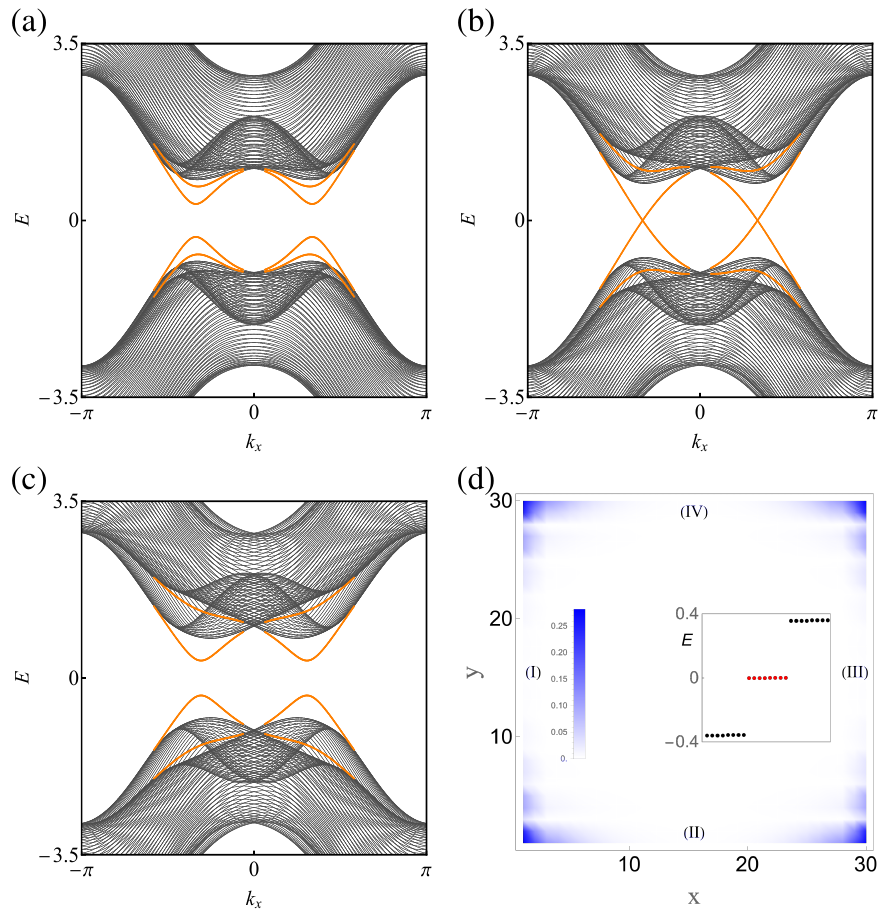}
    \caption{(a-c) The evolution of edge gap on the $y$-normal edges with respect to $\eta$ under a fixed $\Delta_{s}$. (a) $\eta=0.1$. (b) $\eta=0.29$. (c) $\eta=0.5$. (d) Probability density profiles of the eight Majorana
    zero modes (red dots in the inset).
    The inset displays the eigenenergies most close to $E=0$.  (I), (II), (III), and (IV) mark the four edges. The sample size is $30\times 30$. Common parameters are $m=2$, $t_x=t_y=1$, $\lambda=0.5$, $\mu=0$ and $\Delta_s=0.5$.}
    \label{fig.3}
\end{figure}

In Figs.\ref{fig.3}(a-c), we show the evolution
of the edge gap on the $y$-normal edges with respect to $\eta$ under a fixed $\Delta_{s}$. The results show explicitly that the edge gap undergoes a
close-and-reopen transition, which signals
the occurrence of a topological phase transition on the boundary. In the regime
$\eta>\eta_{c}$ ($\eta_{c}$ denotes the critical value at which the edge gap closes),
we further numerically calculate the energy spectra for a sample of square geometry (both $x$ and $y$ directions are cut open). The result shows the existence of eight MZMs whose wave functions are localized around the four corners (second-order boundary),  as shown in Fig. \ref{fig.3}(d).
These numerical results validate the correctness of our analysis and reveal that
the realized SOTSC supports two MZMs per corner.

The existence of twofold MCMs in systems without spinful TRS is nontrivial. Although the Kramers degeneracy dictated by spinful TRS
can prohibit two MZMs at the same corner from coupling\cite{Yan2018hosc,Wang2018hosc}, the lift of spinful TRS will lead to the losing of their stability, unless the existence of protection from an additional symmetry. Here the stability of twofold MCMS is protected by a chiral symmetry.
To be specific, despite the breaking of spinful TRS, the Hamiltonian has an effective spinless TRS ($\mathcal{T}'=\mathcal{K}$). On the other hand, the BdG Hamiltonian is always
particle-hole symmetric, i.e., $\mathcal{C} H(\bm{k}) \mathcal{C}^{-1}=-H(-\bm{k})$ with $\mathcal{C}=\tau_x\mathcal{K}$.
The product of TRS and
PHS gives rise to a chiral symmetry, i.e.,  $\mathcal{S}H(\bk)\mathcal{S}^{-1}= -H(\bk)$
with $\mathcal{S}=\mathcal{T}'\mathcal{C}=\tau_x$. Owing to the existence of chiral symmetry, the MZMs are also eigenstates of the chiral symmetry operator $\mathcal{S}$. The chiral symmetry will
prohibit the MZMs at the same corner from coupling as long as they correspond to the same eigenvalue of
$\mathcal{S}$. This is the underlying
reason why the realized SOTSC can
support twofold MCMs even though the spinful TRS is broken by
the magnetic exchange field.

\subsection{Edge-state theory}
In Sec. \ref{sec: Ham1}, we have demonstrated the existence of twofold MCMs through numerical simulations. In this section, we aim to develop an analytical edge theory to understand the physics.

To simplify the analysis, here we focus on the case with $\mu=0$. All parameters are taken to be positive, with $m<2(t_x+t_y)$ to ensure that the normal state resides within the topological region. Following the standard step\cite{Yan2018hosc}, we first decompose the Hamiltonian Eq. (\ref{eq:TBSH}) into two parts, i.e., $H(\boldsymbol{k})=H_0(\boldsymbol{k})+H_1(\boldsymbol{k})$, with
\begin{equation}
    \begin{aligned}
    H_0(\boldsymbol{k})=\; & (m-2 t_x \cos k_x-2 t_y \cos k_y) \tau_z \sigma_z \\
    &+2 \lambda \sin k_x \sin k_y \tau_z \sigma_y s_y  \,,\\
    H_1(\boldsymbol{k})=\; & 2 \lambda\left(\cos k_x-\cos k_y\right) \tau_z \sigma_x +\Delta_s \tau_y  s_y\\
    & +2 \eta \sin k_x \tau_z \sigma_y s_x \,.
    \end{aligned}
\end{equation}
The coefficient of the first term in $H_0(\boldsymbol{k})$ indicates that band inversion occurs at $(k_x, k_y) = (0, 0)$. Expanding the Hamiltonians around this point to second order in momentum, the continuum counterparts are obtained as
\begin{equation}
    \begin{aligned}
    H_0(k_x, k_y)=\; & (m-2 t_x-2 t_y+t_x k_x^2+t_y k_y^2) \tau_z   \sigma_z \\
                &+2 \lambda k_x k_y \tau_z \sigma_y s_y  \,,\\
    H_1(k_x, k_y)=\; & -\lambda\left(k_x^2-k_y^2\right) \tau_z \sigma_x  +\Delta_s \tau_y  s_y\\
    & +2 \eta k_x \tau_z \sigma_y s_x\,.
    \end{aligned}
    \label{eq:decomposedH}
\end{equation}
Next we are going to determine the low-energy
effective Hamiltonian on the edges of a square
sample with open boundary conditions in $x$ and $y$ directions. For the convenience of discussion, we label the
four edges in an anticlockwise order as I, II, III, and IV, as shown in Fig.\ref{fig.3}(d).

To determine the low-energy effective Hamiltonian on edge I, we assume that the system occupies the whole region with $x\geq0$, so that $x=0$ corresponds to edge I. As the translation symmetry is broken in the $x$ direction, $k_x$ needs to be replaced by the operator $-i \partial_x$ in the Hamiltonian. Accordingly, the two parts of the Hamiltonian become
\begin{align}
    H_0(-i \partial_x,k_y)=\; & \left[M(k_y)-t_x \partial_x^2\right] \tau_z \sigma_z -i v(k_y) \partial_x \tau_z \sigma_y s_y, \nonumber\\
    H_1(-i \partial_x,k_y)=\; & \lambda(\partial_x^2+k_y^2) \tau_z \sigma_x  +\Delta_s \tau_y  s_y -2i \eta \partial_x \tau_z \sigma_y s_x,
\end{align}
where $M(k_y)=m-2 t_x-2 t_y+t_y k_y^2$, $v(k_y)=2 \lambda k_y$. In comparison with
previous studies\cite{Yan2018hosc,Wu2020SOTSC}, here a big difference is that $v(k_y)$ is $k_{y}$-dependent and odd
in $k_{y}$.  Solving the eigenvalue equation $H_0 \psi_\alpha(x)=E_\alpha \psi_\alpha(x)$
under the boundary conditions $\psi_\alpha(0)=\psi_\alpha(+\infty)=0$,
one can find the existence of four zero-energy solutions. The explicit forms of the wave functions are
\begin{equation}
    \psi_\alpha(x)=\mathcal{N}_x \sin (\kappa_1 x) e^{-\kappa_2 x} e^{i k_y y} \chi_\alpha,
\end{equation}
where $\mathcal{N}_x=\sqrt{4\left|\kappa_2(\kappa_1^2+\kappa_2^2) / \kappa_1^2\right|}$ is the normalization constant, $\kappa_1=\sqrt{-\frac{M(k_y)}{t_x}-\frac{v^2(k_y)}{4t_x}}$, $\kappa_2=\frac{\xi v(k_y)}{2t_x}$,
and the spinor $\chi_\alpha$ represents an eigenstate of the operator $\sigma_x s_y$, i.e., $\sigma_x s_y\chi_\alpha=\xi\chi_\alpha$
with $\xi=\pm1$.
As the boundary conditions require $\kappa_2>0$ and $\kappa_2$ is proportional
to $\xi v(k_y)$, $\chi_{\alpha}$ turns out to sensitively dependent on $k_{y}$. To be specific, when $k_y>0$, $v(k_{y})>0$, so
$\xi=1$, i.e., $\sigma_x s_y\chi_\alpha=\chi_\alpha$. The corresponding four eigenvectors can be chosen as
\begin{equation}
    \begin{aligned}
    & \chi_1^+=\;\ket{\tau_z=+1}\otimes\ket{\sigma_x=+1}\otimes\ket{s_y=+1}\,, \\
    & \chi_2^+=\;\ket{\tau_z=+1}\otimes\ket{\sigma_x=-1}\otimes\ket{s_y=-1}\,, \\
    & \chi_3^+=\;\ket{\tau_z=-1}\otimes\ket{\sigma_x=+1}\otimes\ket{s_y=+1}\,, \\
    & \chi_4^+=\;\ket{\tau_z=-1}\otimes\ket{\sigma_x=-1}\otimes\ket{s_y=-1}\,.
    \end{aligned}
\end{equation}
When $k_y<0$, $v(k_{y})<0$, so $\xi=-1$, i.e., $\sigma_x s_y\chi_\alpha=-\chi_\alpha$. The corresponding four eigenvectors can be chosen as
\begin{equation}
    \begin{aligned}
    & \chi_1^-=\;\ket{\tau_z=+1}\otimes\ket{\sigma_x=-1}\otimes\ket{s_y=+1}\,, \\
    & \chi_2^-=\;\ket{\tau_z=+1}\otimes\ket{\sigma_x=+1}\otimes\ket{s_y=-1}\,, \\
    & \chi_3^-=\;\ket{\tau_z=-1}\otimes\ket{\sigma_x=-1}\otimes\ket{s_y=+1}\,, \\
    & \chi_4^-=\;\ket{\tau_z=-1}\otimes\ket{\sigma_x=+1}\otimes\ket{s_y=-1}\,.
    \end{aligned}
\end{equation}
By projecting $H_1$ onto the subspace spanned by $\psi_\alpha(x)$, one obtains the low-energy effective Hamiltonian on edge I, whose matrix elements are given by
\begin{equation}
    H_{\mathrm{I}, \alpha \beta}(k_y)=\;\int_0^{+\infty} d x \psi_\alpha^*(x) H_1(-i \partial_x, k_y) \psi_\beta(x)\,.
\end{equation}
After some simple calculations, one can find
\begin{equation}\label{EH1}
    H_{\rm I} (k_y)=\left\{
        \begin{aligned}
        \lambda\left(\frac{M(k_y)}{t_x}+k_y^2\right)\tau_z s_z + \Delta_s\tau_y s_z,\,\,&k_y>0,\\
        -\lambda\left(\frac{M(k_y)}{t_x}+k_y^2\right)\tau_z s_z + \Delta_s\tau_y s_z,\,\,&k_y<0.
        \end{aligned}
    \right.
\end{equation}
The existence of two pairs of helical edge states in the normal state leads to a novel characteristic in the effective Hamiltonian: the presence of two valleys, where the valley momenta are determined by the condition that the coefficient of the first term in Eq.(\ref{EH1}) vanishes. The valley momenta can be readily identified as $K_\pm=\pm k_{0}$, where $k_{0}=\sqrt{{(2t_x+2t_y-m)}/{(t_x+t_y)}}$. Labeling the valley with positive (negative) momentum as $+$ ($-$), the effective Hamiltonian can be expanded around the two valleys to first order in momentum, yielding the following valley Hamiltonians
\begin{equation}
    \begin{aligned}
        H_{\rm  I}^+(q_y) =\; \Lambda_{\rm  I} q_y\tau_z s_z + \Delta_s\tau_y s_z \,,\\
        H_{\rm  I}^-(q_y) =\; \Lambda_{\rm  I} q_y\tau_z s_z + \Delta_s\tau_y s_z \,,
    \end{aligned}
\end{equation}
where $q_{y}$ denotes a small momentum measured from the corresponding valley momentum, and $\Lambda_{\rm  I}=2\lambda(t_x+t_y)k_0/t_x$. The above
valley Hamiltonians take the standard form
of Dirac Hamiltonians, with the first term
denoting the kinetic energy and the second term denoting the Dirac mass.

Similarly, the low-energy effective Hamiltonians for the other three edges can be readily derived. For the positive valley, the low-energy edge Hamiltonians are given by
\begin{equation}
    \begin{aligned}
        H_{\rm  I}^+(q_y) &= \Lambda_{\rm  I} q_y\tau_z s_z + \Delta_s\tau_y s_z \,,\\
        H_{\rm  II}^+(q_x) &= -\Lambda_{\rm  II} q_x\tau_z s_z + \Delta_s\tau_y s_z - 2\eta k_0\tau_z s_y \,,\\
        H_{\mathrm{III}}^{+}(q_y)&=-\Lambda_{\rm  III} q_y \tau_z s_z+\Delta_s \tau_y s_z \,,\\
        H_{\rm  IV}^+(q_x) &= \Lambda_{\rm  IV} q_x\tau_z s_z + \Delta_s\tau_y s_z + 2\eta k_0\tau_z s_y \,.
    \end{aligned}
    \label{eq:fourHe}
\end{equation}
The low-energy edge Hamiltonians for the negative valley, $H_\alpha^-$, have exactly the same form as $H_\alpha^+$, where the edge index $\alpha\in\{\rm{I,II,III,IV}\}$. The low-energy edge Hamiltonians reveal that the $s$-wave pairing term generates a uniform Dirac mass on all edges, while the $\eta$ term contributes to the Dirac mass only on edges II and IV. This conclusion is consistent with the numerical results presented in Sec. \ref{sec: Ham1}. Viewing the boundary as a 1D periodic system and introducing a coordinate $l$ to characterize it\cite{Yan2018hosc}, the low-energy edge Hamiltonians in Eq. (\ref{eq:fourHe}) can be expressed in a unified form
\begin{equation}
H_{\text {edge }}^+=i \Lambda (l) \tau_z s_z \partial_l+\Delta_s \tau_y s_z + M_\eta(l) \tau_z s_y\,,
\end{equation}
where $\Lambda(l)=2\lambda(t_x+t_y)k_0/t_x,\,2\lambda(t_x+t_y)k_0/t_y,\,2\lambda(t_x+t_y)k_0/t_x,\,2\lambda(t_x+t_y)k_0/t_y$ and $M_\eta(l)=0,\,-2\eta k_0,\,0,\,2\eta k_0$ from edge I to edge IV. Again, the Hamiltonian $H_{\text {edge }}^-$ has the same form as $H_{\text {edge }}^+$.

As $H_{\text {edge }}^-$ and $H_{\text {edge }}^+$ have the same form, below we only consider  $H_{\text {edge }}^+$ for discussion. Apparently, the signs of $\Lambda(l)$ on adjacent edges are the same, whereas the Dirac masses on adjacent edges can be tuned to take opposite signs by adjusting the parameters $\eta$ and $\Delta_s$. When
the signs of the Dirac masses on two adjacent edges are opposite, a Dirac-mass domain wall is formed at the corner, resulting in the presence of a MZM at the corner.
Let us first focus on the intersection of edge I and edge II.  A zero-energy eigenstate $\psi(l)$ will satisfy the eigenvalue equation $H_{\text {edge }}^+\psi(l)=0$.  The solution will take the generic form
\begin{equation}
\psi(l) \propto e^{\int^l d l^{\prime} \left.\left(\alpha\Delta_s+\beta M_\eta(l^{\prime})\right) \right/ \Lambda(l^{\prime})}\left|\tau_x=\alpha,s_x=\beta\right\rangle \,.
\end{equation}
Under the boundary conditions $\psi(-\infty)=\psi(\infty)=0$, a Majorna zero mode localized around the corner exists when $\alpha=1,\beta=1$, and the parameter condition $\Delta_s-2\eta k_0<0$ is satisfied. It is straightforward to verify that, when the above parameter condition is satisfied, MZMs also exist at the other three corners. Because $H_{\text {edge }}^-$ and $H_{\text {edge }}^+$ give rise to the same physics,  each corner hence supports two MZMs which are the eigenstates of the chiral symmetry operator corresponding to the same eigenvalue. Up to a normalization constant, the wave functions for these MCMs can be expressed as
\begin{equation}
    \begin{aligned}
      \psi(l)_1^\pm &\propto e^{\int^l d l^{\prime} \left. \left( \Delta_s + M_p(l^{\prime})\right) \right/ \Lambda(l^{\prime})}\left|\tau_x=+1,s_z=+1\right\rangle\,,  \\
      \psi(l)_2^\pm &\propto e^{\int^l d l^{\prime} \left.-\left(\Delta_s+ M_p(l^{\prime})\right) \right/ \Lambda(l^{\prime})}\left|\tau_x=-1,s_z=-1\right\rangle\,,\\
      \psi(l)_3^\pm &\propto e^{\int^l d l^{\prime} \left. \left( \Delta_s- M_p(l^{\prime})\right) \right/ \Lambda(l^{\prime})}\left|\tau_x=+1,s_z=-1\right\rangle\,,  \\
      \psi(l)_4^\pm &\propto e^{\int^l d l^{\prime} \left.-\left(\Delta_s- M_p(l^{\prime})\right) \right/ \Lambda(l^{\prime})}\left|\tau_x=-1,s_z=+1\right\rangle\,.
    \end{aligned}
    \label{cornermodes}
\end{equation}
To explicitly determine the chirality of the MCMs, one needs to  project
the chiral symmetry operator onto the subspace spanned by the basis states of the low-energy edge Hamiltonian. It turns out that its projected form  is (here we restore the identity matrices in the orbital ($\sigma_0$) and spin ($s_{0}$) subspaces to see the effect of the projection)
\begin{equation}
    \mathcal{S}:\tau_x\sigma_0 s_0\to\tau_x s_0\,.
\end{equation}
According to the projected form, it is evident that the chiralities of the MCMs in Eq. (\ref{cornermodes}) are $(++, --, ++, --)$ in order, which indicates that the two MZMs at the same corner share the same chirality, verifying that their robustness is protected by chiral symmetry.

Based on the edge-state theory,
the topological criterion for the presence of twofold MCMs at $\mu=0$ is determined as
\begin{equation}
    0<\Delta_s<2\eta k_{0}.
    \label{eq:TRB&Scriterion}
\end{equation}

\subsection{Bulk topological invariant }
As discussed in Sec.~\ref{sec: Ham1}, the Hamiltonian presented in Eq.~(\ref{eq:TBSH}) has chiral symmetry. In  Ref.\cite{Benalcazar2022CSHOTI}, Benalcazar and Cerjan showed that the number of zero-energy corner states in a chiral symmetric system can be characterized by the so-called multipole chiral number $N_{xy}$. Interestingly, the multipole chiral number is defined under periodic boundary conditions, and therefore it is a bulk topological invariant.

\begin{figure}[]
    \includegraphics[width=0.7\columnwidth]{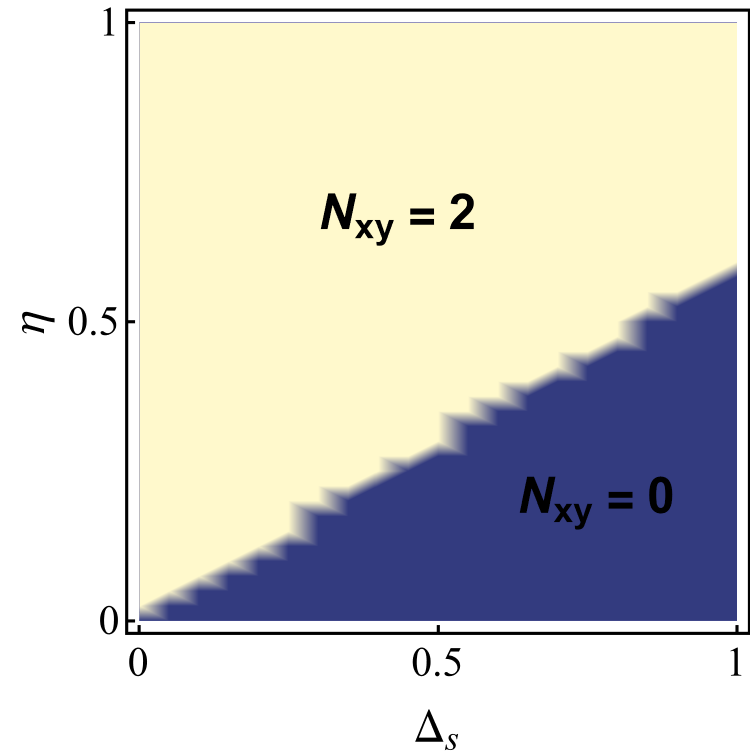}
    \caption[]{The phase diagram of $N_{xy}$ with respect to $\Delta_s$ and $\eta$ is depicted, where the light yellow region represents the topologically non-trivial phase, indicating the presence of twofold Majorana corner states. The remaining parameters are set to $m = 2$, $t_x = t_y = 1$, $\mu = 0$, $\lambda = 0.5$, and the lattice size $L_x \times L_y = 30 \times 30$.} \label{fig.4}
\end{figure}

To illustrate the bulk-corner correspondence,
we again focus on $\mu=0$ and numerically calculate the multipole chiral number for a system with size $L_x \times L_y = 30 \times 30$ and periodic boundary conditions in both $x$ and $y$ directions (more details are provided in Appendix \ref{sectin:appC}).
A typical phase diagram in accordance with
the multipole chiral number is shown in
Fig.\ref{fig.4}. We find
that the phase boundary agrees very well with
the one determined by the edge-state theory (Eq.(\ref{eq:TRB&Scriterion})), and
indeed $N_{xy}=2$ is found for the set of parameters
considered in Fig.\ref{fig.3}(d), demonstrating that this bulk topological invariant can predict the number of MZMs per corner.

\subsection{Effects of the chemical potential}

As the chiral symmetry is preserved even when the chemical potential is tuned away from $\mu=0$, the MZMs will be
robust for a range of $\mu$. Empirically,
the variation of chemical potential may induce
the close of energy gap in the bulk or at the boundary, leading to the change of topology
accordingly. Let us first investigate its  influence on the bulk spectra. According to the Hamiltonian (\ref{eq:TBSH}), the bulk spectra are given by
\begin{equation}
    E(\bk)=\pm\sqrt{F(\bk) \pm 2\sqrt{
    G(\bk)
    }}.\nonumber
\end{equation}
where $F(\bk)=\epsilon^2(\bk)+\mu^2+\eta^2(\bk)+\Delta_s^2$ and $G(\bk)=\mu^2(\epsilon^2(\bk)+\eta^2(\bk))+\eta^2(\bk)\Delta_s^2$, with $\eta(\bk)=2\eta \sin k_x$ and
\begin{equation}
    \begin{aligned}
        \epsilon(\bk)=&\{\left(m-2 t_x \cos k_x-2 t_y \sin k_y \right)^2+\left( 2 \lambda \sin k_x \sin k_y\right)^2\\
         &+\left[2 \lambda\left(\cos k_x-\cos k_y\right)\right]^2\}^{1/2}.\nonumber
    \end{aligned}
\end{equation}
It is easy to check that the $s$-wave pairing term introduces a bulk gap that remains opens under the variation of all other parameters,
including $\mu$. This indicates that the first-order topology will keep trivial with the change of $\mu$. However, as the system must be topologically trivial for all orders when $\mu$ goes to infinity [all states are occupied ($\mu=+\infty$) or empty ($\mu=-\infty$)], the edge gap must undergo close-and-reopen transitions with the increase of $|\mu|$, leading to topological phase transitions from the SOTSC to a trivial superconductor at last.

\begin{figure}[]
    \centering
    \includegraphics[width=0.96\columnwidth]{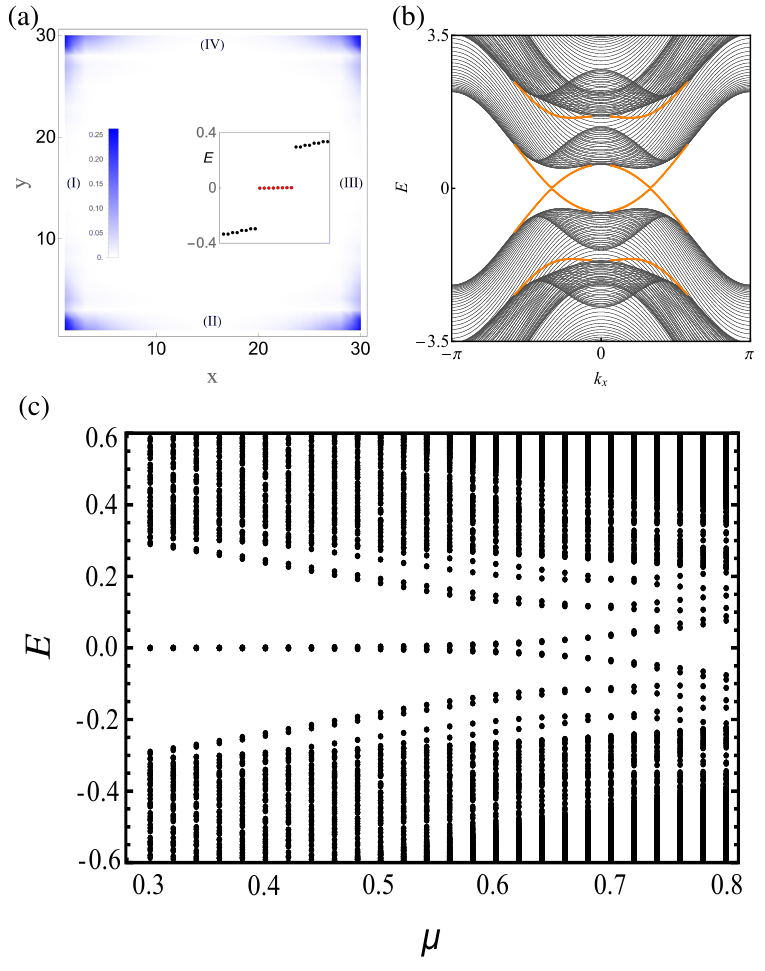}
    \caption{(a) The probability density profiles of Majorana corner modes and energy spectra of a square geometry with $\mu=0.3$, showing that each corner continues to host two Majorana zero modes.  (b) Energy spectra for a cylinder geometry with open boundary conditions in the $y$ direction at $\mu=0.68$, where the boundary energy gap closes, indicating the critical point of a boundary topological phase transition.  (c) The evolution of energy spectra as a function of $\mu$ for a square geometry, illustrating that small values of $\mu$ do not alter the system's topological characteristics. Common parameters are $m=2$, $t_x=t_y=1$, $\lambda=0.5$ and $\Delta_s=\eta=0.5$. }
    \label{fig.5}
\end{figure}

As the normal-state Hamiltonian has PHS, the topology of the BdG Hamiltonian will be the same for opposite $\mu$. Therefore, an investigation of the region with positive $\mu$ is sufficient to fully understand the system. By fixing
all parameters except $\mu$ to be the same as in Fig.\ref{fig.3}(a) and only varying $\mu$,
we find that the twofold MCMs are stable when $\mu<\mu_{c}=0.68$, as shown in Fig.\ref{fig.5}(a). Here $\mu_{c}$
corresponds to the critical value that the edge gap on $y$-normal edges gets closed, as
shown in Fig.\ref{fig.5}(b). When $\mu>\eta_{c}$, all MCMs disappear (see Fig.\ref{fig.5}(c)), suggesting that the system enters a topologically trivial phase.
The results confirm that the twofold MCMs are stable for a wide range of $\mu$, and the variation of $\mu$ can induce a topological phase transition at the boundary.

Before ending this section, we remark that, when the chemical potential is nonzero, even when $\mu$ exceeds $\mu_{c}$, the multipole chiral number calculated is still found to take the nontrivial value $2$, suggesting its failure to characterize the topological properties of the system when $\mu$ deviates
away from zero. How to define a bulk topological invariant to  accurately describe the system's topology under all kinds of conditions remains an open question for future research.

\section{Second-order topological superconductors with twofold Majorana corner modes based on mixed-parity superconductivity}\label{IV}

In the previous section, we have showed
that a SOTSC supporting twofold MCMs
can be achieved when the two pairs of
helical edge states of the quantum spin Hall insulator are gapped by the $s$-wave pairing and magnetic exchange field.
In this section, we are going to show that
similar physics can be obtained if the
helical edge states are gapped by a mixed-parity pairing consisting of $s$-wave and $p$-wave pairing components.

\subsection{Bulk Hamiltonian}
The BdG Hamiltonian to concern is of the form
\begin{align}\label{eq:SPmodel}
    H(\boldsymbol{k}) =\; &(m-2 t_x \cos k_x-2 t_y \cos k_y) \tau_z \sigma_z  - \mu \tau_z \nonumber\\
    &+ 2 \lambda(\cos k_x-\cos k_y) \tau_z \sigma_x  \nonumber\\
    &+ 2 \lambda \sin k_x \sin k_y \tau_z \sigma_y s_y \\
    &+ 2 \Delta_p \sin k_x \tau_y + \Delta_s \tau_ys_y \,.\nonumber
\end{align}
Compared to Eq.(\ref{eq:TBSH}), the only change is a replacement of the magnetic exchange field by an odd-parity $p$-wave pairing component, the $\Delta_{p}$ term.
The above Hamiltonian may describe a heterostructure composed of the quantum spin Hall insulator and an $s$-wave superconductor,
where the inversion symmetry is broken, and hence the coexistence of even-parity and odd-parity pairing components is permitted\cite{Gorkov2001mixed}.

When concerning the topology of a Hamiltonian,
the algebraic commutation and anticommutation relation of its terms are essential\cite{Ryu2010}. In comparison with the Hamiltonian in Eq.~(\ref{eq:TBSH}),
one can easily find that, at the limit $\mu=0$, the algebraic relations among the terms of the Hamiltonian in  Eq.~(\ref{eq:SPmodel}) are the same, e.g., the
$p$-wave pairing term also commutes with the
$s$-wave pairing terms, while anticommuting with all other terms. Based on this simple fact, one can expect that this Hamiltonian can
aslo realize SOTSCs.

Despite the aforementioned similarity in the algebraic structure, the $p$-wave pairing term has the following two important differences when compared with the magnetic exchange field. First, the $p$-wave pairing term anticommutes with the chemical potential. Second, it does not break the spinful TRS ($\mathcal{T} = i s_y \mathcal{K}$ and $\mathcal{T}^{2}=-1$). Therefore, the BdG Hamiltonian
in  Eq.~(\ref{eq:SPmodel}) has a natural chiral symmetry which is the product of the spinful TRS and PHS ($\mathcal{C} = \tau_x\mathcal{K}$). The chiral symmetry operator is of the form $\mathcal{S}=-i\mathcal{T} \mathcal{C}=\tau_{x}s_{y}$.

In a system with spinful TRS, the MZMs will always
show up in pairs due to the Kramers degeneracy. As $\mathcal{C}\mathcal{T}=-\mathcal{T}\mathcal{C}$, the Majorana Kramers pair will be the eigenstates of the chiral symmetry operator corresponding to opposite eigenvalues. To see this, consider $|\psi_{1}\rangle$ to be the wave function of one MZM, and  $|\psi_{2}\rangle=\mathcal{T}|\psi_{1}\rangle$
refers to the wave function of its time-reversal partner. As  both $|\psi_{1}\rangle$
and $|\psi_{1}\rangle$ are eigenstates of $\mathcal{C}$, if $\mathcal{C}|\psi\rangle_{1}=\xi|\psi\rangle_{1}$ ($\xi=\pm1$), it is easy to find  $\mathcal{C}|\psi_{2}\rangle=\mathcal{C}\mathcal{T}|\psi_{1}\rangle=-\mathcal{T}\mathcal{C}|\psi_{1}\rangle=-\xi\mathcal{T}|\psi_{1}\rangle=-\xi|\psi_{2}\rangle$. This is an important property which implies that the chiral symmetry cannot provide additional protection to Majorana Kramers pairs. In other words, in a spinful time-reversal invariant SOTSC, if there is no additional crystalline symmetry or symmetry involving internal degrees of freedom, each corner can at most host two robust MZMs.
This is distinct from the scenario in a  SOTSC with spinless TRS where an arbitrary number of MZMs can be stabilized at one corner.

Interestingly, the Hamiltonian in Eq.~(\ref{eq:SPmodel}) has a $U(1)$ rotation symmetry in the spin space. The existence
of this symmetry can give rise to a new set
of TRS operator and chiral symmetry operator. To be specific, because
$[s_{y},H(\bk)]=0$, one can check $\mathcal{T'}H(\bk)\mathcal{T'}^{-1}=H(-\bk)$
and $\mathcal{S'}H(\bk)\mathcal{S'}^{-1}=-H(\bk)$, where $\mathcal{T'}=-is_{y}\mathcal{T}=\mathcal{K}$, and $\mathcal{S'}=s_{y}\mathcal{S}=\tau_{x}$. It is apparent that $\mathcal{T'}^{2}=1$, indicating that
the Hamiltonian has an effective spinless TRS. In addition, it is also easy to check
$\mathcal{T'}\mathcal{S'}=\mathcal{S'}\mathcal{T'}$ and $\mathcal{T}\mathcal{S'}=\mathcal{S'}\mathcal{T}$. This commutative property indicates that this chiral symmetry can protect the existence of multifold MCMs.

\subsection{Edge-state theory}

\begin{figure}
    \centering
    \includegraphics[width=0.6\linewidth]{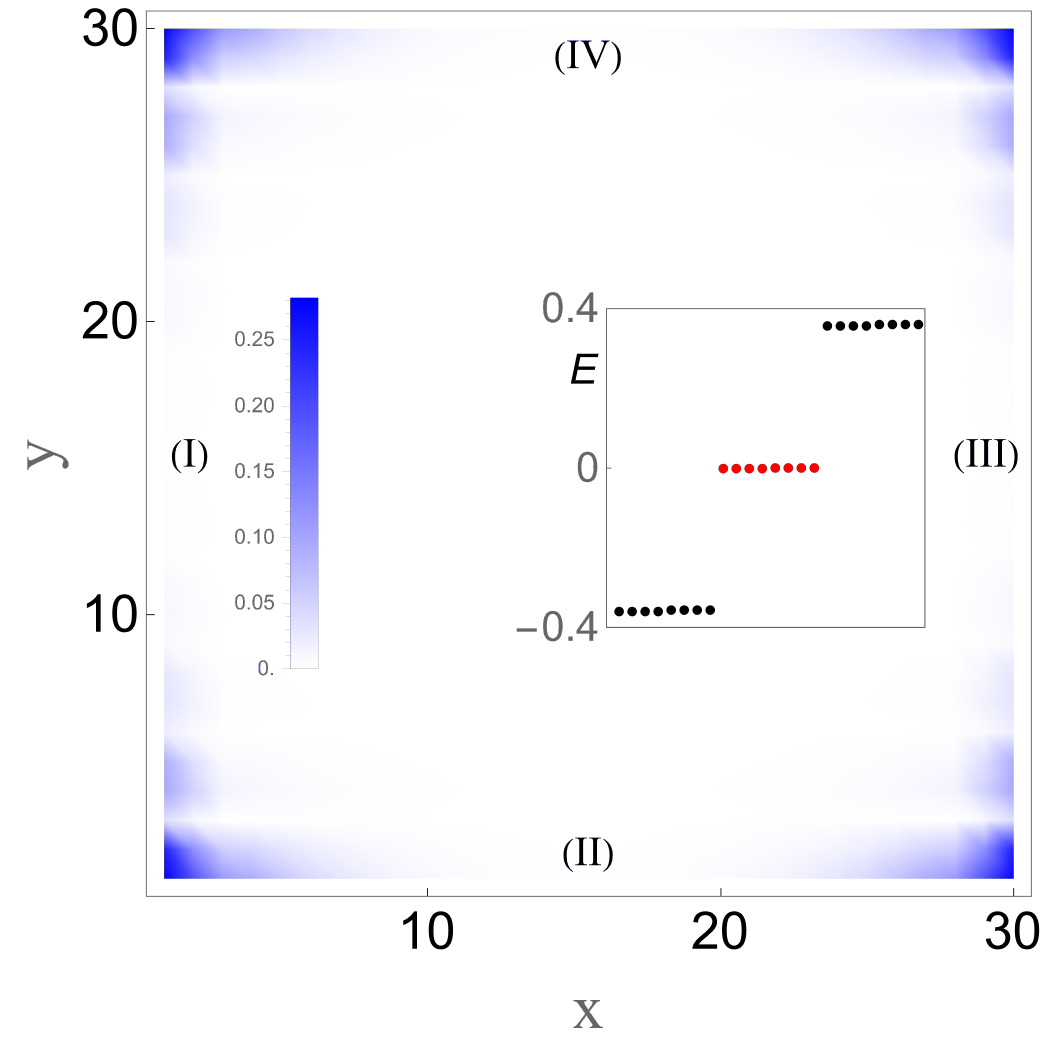}
    \caption{The probability density profiles of the Majorana corner modes. The inset illustrates the energy spectrum near $E=0$, revealing the presence of eight Majorana zero modes. The parameters are set to $m=2, t_x=t_y=1, \lambda=0.5, \Delta_s=0.5, \Delta_p=0.5, \mu=0$, and the lattice size of the model is $30 \times 30$.}
    \label{fig:SPcornermode}
\end{figure}

To have a more intuitive understanding
of the similarity in topology between the two BdG Hamiltonian, in this part we also consider $\mu=0$ and develop the edge-state theory for the second BdG Hamiltonian.
Applying the same convention to the labels of the four edges and following the same steps(detailed in Appendix \ref{sectin:appB}),
we find that the low-energy edge Hamiltonians
take a similar form,
\begin{equation}
    \begin{aligned}
        H_{\text{edge } \alpha}^+ &= i \Lambda_{\alpha} \tau_z s_z \partial_l + \Delta_s \tau_y s_z + M_p(\alpha) \tau_y, \\
        H_{\text{edge } \alpha}^- &= i \Lambda_{\alpha} \tau_z s_z \partial_l + \Delta_s \tau_y s_z - M_p(\alpha) \tau_y,
    \end{aligned}
\end{equation}
where $\alpha=\{\text{I}, \text{II}, \text{III}, \text{IV}\}$,
\begin{align}
    \Lambda_{\rm I} & = \Lambda_{\rm III} = 2\lambda(t_x + t_y)k_0 / t_x , \notag \\
    \Lambda_{\rm II} & = \Lambda_{\rm IV} = 2\lambda(t_x + t_y)k_0 / t_y , \notag \\
    M_p(\rm I) & = M_p(\rm{III}) = 0 ,\, M_p(\rm{II}) = M_p(\rm{IV}) = 2\Delta_p k_0. \notag
\end{align}
Here $M_p(\alpha)$ denotes the Dirac mass induced by the $p$-wave pairing term at the $\alpha$ edge
From the edge dependence of $M_{p}$, it is easy to see that the effects of
the  $p$-wave pairing term and the magnetic exchange field on the helical edge states are quite similar. Immediately, one knows that
the criterion for the presence of twofold MCMs at $\mu=0$ is
\begin{equation}
    0<\Delta_s < 2\Delta_p k_0.
    \label{eq:SPcriterion}
\end{equation}
The numerical result in Fig.~\ref{fig:SPcornermode} validates the existence of twofold MCMs when the above topological criterion is
satisfied, confirming that this Hamiltonian
can also realize a SOTSC with twofold MCMs.

Also based on the Jackiw-Rebbi theory, the wave functions of the eight MZMs at the four corners are given by
\begin{equation}
    \begin{aligned}
      \psi(l)_1^\pm &\propto e^{\int^l d l^{\prime} \left. \left( \Delta_s -M_p(l^{\prime})\right) \right/ \Lambda(l^{\prime})}\ket{\tau_x=+1,s_z=\mp 1}\,,  \\
      \psi(l)_2^\pm &\propto e^{\int^l d l^{\prime} \left.-\left(\Delta_s- M_p(l^{\prime})\right) \right/ \Lambda(l^{\prime})}\ket{\tau_x=-1,s_z=\mp 1}\,,\\
      \psi(l)_3^\pm &\propto e^{\int^l d l^{\prime} \left. \left( \Delta_s- M_p(l^{\prime})\right) \right/ \Lambda(l^{\prime})}\ket{\tau_x=+1,s_z=\mp 1}\,,  \\
      \psi(l)_4^\pm &\propto e^{\int^l d l^{\prime} \left.-\left(\Delta_s- M_p(l^{\prime})\right) \right/ \Lambda(l^{\prime})}\ket{\tau_x=-1,s_z=\mp 1}\,.
    \end{aligned}
    \label{eq:SPcornermode}
\end{equation}
According to the wave functions of the MZMs at the same corner, one can see that $\psi(l)_{\beta}^{+}=\mathcal{T}\psi(l)_{\beta}^{-}$ ($\beta=\{1,2,3,4\}$), which is a natural consequence of the spinful TRS.
On the other hand, one can also find that $\mathcal{S'}\psi(l)_{\beta}^{\pm}=\xi\psi(l)_{\beta}^{\pm}$, namely, the wave functions of the two MZMs at the same corner correspond to the same eigenvalue of the chiral symmetry operator $\mathcal{S'}$. This means that the robustness of the twofold MCMs can be attributed either to  the spinful TRS or to the chiral symmetry
$\mathcal{S'}$. Indeed, we find that introducing a weak Zeeman field of the form $B_{x}\tau_{z}\sigma_{0}s_{x}+B_{z}\tau_{z}\sigma_{0}s_{z}$, which breaks the spinful TRS but preserves the chiral symmetry $\mathcal{S'}$, the twofold MCMs remain robust, demonstrating that the chiral symmetry $\mathcal{S'}$ only is sufficient to protect the existence of multifold MCMs.

\subsection{Bulk topological invariant}

\begin{figure}[t]
    \centering
    \includegraphics[scale=0.5]{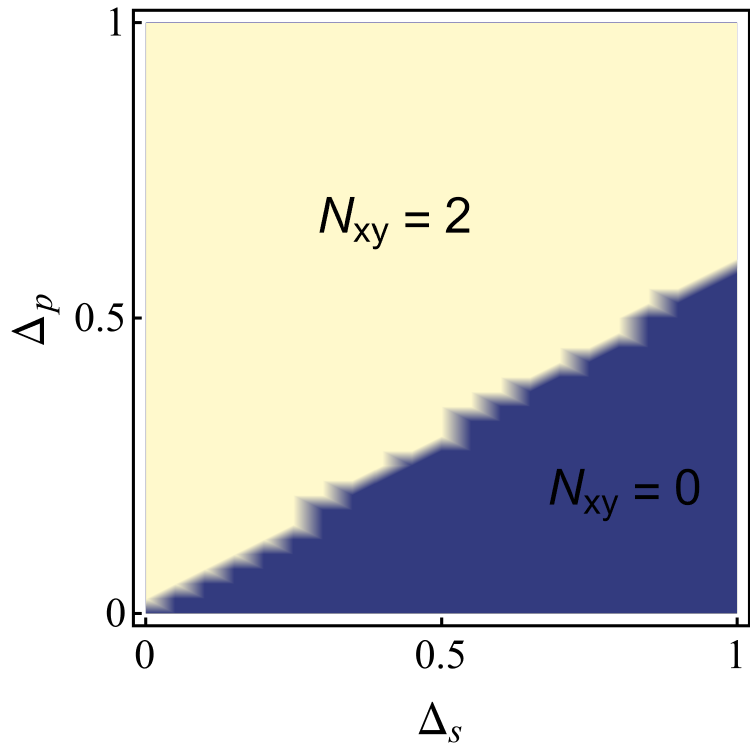}
    \caption{Phase diagram determined by the multipole chiral number $N_{xy}$ as a function of the variables $(\Delta_s, \Delta_p)$. Other parameters are set to $m = 2$, $t_x = t_y = 1$, $\mu = 0$, $\lambda = 0.5$, and the lattice size $L_x \times L_y = 30 \times 30$.}
    \label{fig.7}
\end{figure}

As aforementioned, owing to the existence of an internal symmetry in the spin space, the BdG Hamiltonian in Eq.(\ref{eq:SPmodel}) has two types of chiral symmetry, $\mathcal{S}$ and $\mathcal{S'}$. Therefore, in principle two multipole chiral numbers can be defined accordingly. However, as $\mathcal{T}\mathcal{S}=-\mathcal{S}\mathcal{T}$, the multipole chiral number associated with $\mathcal{S}$ will always be trivial. Therefore, we will focus on the multipole chiral number defined based  on $\mathcal{S'}$. A representative phase diagram at $\mu=0$ is shown in Fig.\ref{fig.7}. In the region with $N_{xy}=2$, the system is a SOTSC with twofold MCMs. The phase boundary is nearly given by $\Delta_s \approx 2\Delta_p$ for the set of parameters we choose, which agrees very well with the criterion
$\Delta_s \approx 2\Delta_p k_{0}$ obtained by edge-state theory.

One can see that the phase diagram is in fact the same as the one presented in Fig.\ref{fig.4}. This is not accidental.
As we pointed out, when $\mu=0$, the commutation and anticommutation relations among the terms in the two BdG Hamiltonians are the same,
so their phase diagrams are also the same.

\subsection{Effects of the chemical potential}

When $\mu\neq0$, as the $p$-wave pairing term is anticommutative with the chemical potential, which is distinct from the magnetic exchange field which is commutative with the chemical potential, one can expect that the effects of the chemical potential on the two BdG Hamiltonians will display some differences.

\begin{figure}[t]
    \centering
    \includegraphics[width=0.5\linewidth]{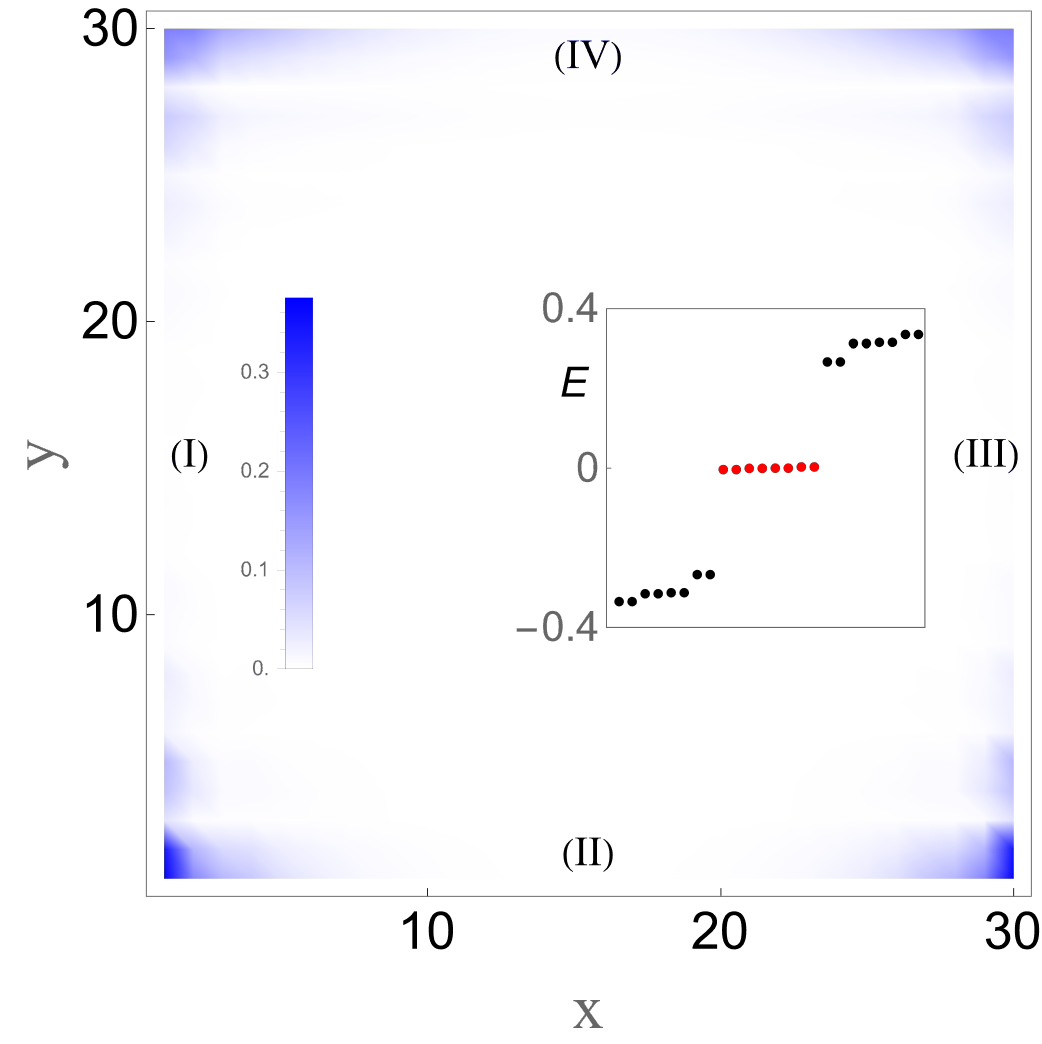}
    \caption{Probability density profiles of the Majorana corner modes for $\mu=0.3$. Also two Majorana zero modes per corner are found. The inset shows the energy spectrum near $E = 0$. We set  $\mu=0.3, m=2, t_x=t_y=1, \lambda=0.5, \Delta_s=0.5, \Delta_p=0.5$, and the lattice size is $30 \times 30$.}
    \label{fig:SPmu}
\end{figure}

When $\mu$ is much smaller than the energy scale of the bulk gap and edge gap, its effect to the MCMs can be determined through the low-energy edge Hamiltonian. To be specific, when $\mu$ is small but finite, the edge Hamiltonian will be modified as
\begin{equation}
    \begin{aligned}
        H_{\text {edge }}^+ &= i \Lambda (l) \tau_z s_z \partial_l + \Delta_s \tau_y s_z + M_p(l) \tau_y - \mu \tau_z, \\
        H_{\text {edge }}^- &= i \Lambda (l) \tau_z s_z \partial_l + \Delta_s \tau_y s_z - M_p(l) \tau_y - \mu \tau_z.
    \end{aligned}
\end{equation}
Without loss of generality, let us focus on  the corner at the intersection of edge I and edge II.
For the MCM originating from $H_{\text {edge }}^+$,
it is easy to find that its wave function becomes
\begin{equation}
    \psi(l) \propto e^{\int^l d l' \left[\Delta_s - M_p(l') +  i \mu\right] / \Lambda(l')} \left|\tau_x=+1,s_z=-1\right\rangle \,.
\end{equation}
Similarly, the wave function of the MCM originating from  $H_{\text {edge }}^-$ can be written as
\begin{equation}
    \psi(l) \propto e^{\int^l d l' \left[\Delta_s - M_p(l') -  i \mu\right] / \Lambda(l')} \left|\tau_x=+1,s_z=+1\right\rangle \,.
\end{equation}
The wave fucntions of the MZMs at the other three corners have a similar form. The above results imply that the chemical potential only adds a phase factor into the wave functions and has little effect on the density profile of the MCMs.
Therefore, the MCMs are also expected to be stable against
the change of chemical potential as long as the variation does not induce
a gap close on the edge or in the bulk.
In Fig.~\ref{fig:SPmu}, the result show that the twofold MCMs remain
well-localized when $\mu$ is increased from $0$ to $0.3$.

\begin{figure}[t]
    \centering
    \includegraphics[width=0.96\columnwidth]{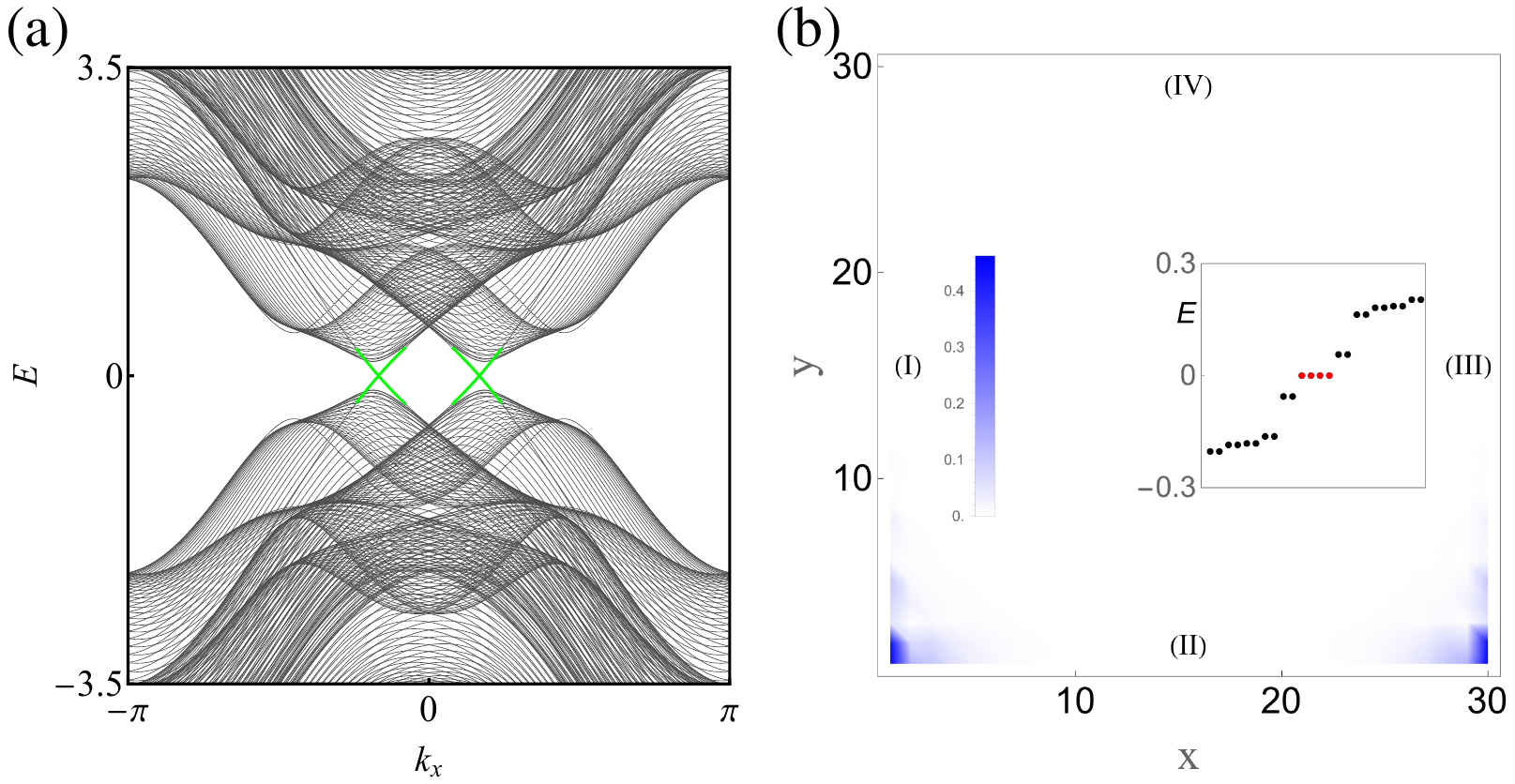}
    \caption{As the chemical potential continues to increase, a topological phase transition will occur at the boundary. (a) At the critical value $\mu=0.65$, the energy spectrum with open boundary conditions along the $y$ direction, where the edge gap disappears. (b) The spatial distribution of Majorana corner modes when $\mu=0.65$, where the corner states on the upper boundary disappear, and the energy spectrum in the inset indicates the existence of only four Majorana zero modes. Here, $m=2, t_x=t_y=1, \lambda=0.5, \Delta_s=0.5, \Delta_p=0.5$.}
    \label{fig:SPcritical}
\end{figure}

As the chemical potential continues to increase, a close-and-reopen transition of the edge gap similar to the previous BdG Hamiltonian is also expected. Through numerical calculations, we find that this is true, however, a notable difference is also found. Owing to the breaking of  inversion symmetry and mirror symmetry $\mathcal{M}_{y}$  by the mixed-parity pairing, we find that the effects of the chemical potential on the energy gap of the top and bottom $y$-normal edges are different. As shown in Fig.~\ref{fig:SPcritical}(a),
the energy gap on the top $y$-normal edge gets closed at a critical chemical
potential. Across the critical point,
the four MZMs at the top
two corners disappear, while the four at the bottom two corners are intact, as shown in Fig.~\ref{fig:SPcritical}(b). This result indicates that now the top $y$-normal edge becomes topologically equivalent with the left and right $x$-normal edges but topologically distinct with the bottom $y$-normal edges.

With a further increase of the chemical potential, the system can become a Dirac superconductor with nodal points in its bulk energy bands. This is another remarkable difference compared with the previous BdG Hamiltonian which always has a gapped bulk energy spectrum. To see the possibility of the realization of a Dirac superconductor, we explicitly write down the bulk energy spectrum of the Hamiltonian  (\ref{eq:SPmodel}), which reads
\begin{equation}
    E(\bm{k}) = \pm \sqrt{\left(\epsilon(\bm{k}) \pm \mu\right)^2 + \left(\Delta_s \pm 2 \Delta_p \sin k_x\right)^2}.
\end{equation}
Apparently, when the zero contours satisfying $\epsilon(\bm{k})-\mu=0$
and $\Delta_s \pm 2 \Delta_p \sin k_x=0$
cross, the energy spectrum has nodal points. Interestingly, we find that the MCMs can remain well-localized when the system just undergoes the transition from a gapped superconductor to a Dirac superconductor. However, a further increase of the chemical potential will eventually make all MCMs disappear.

\section{Discussions and conclusions}\label{V}

We have shown that quantum spin Hall insulators with two pairs of helical egde states provide a basis to realize SOTSCs with twofold MCMs even when the spinful TRS is broken, provided that the chiral symmetry is preserved.
This correspondence between the spin Chern number and the number of MZMs per corner can be straightforwardly generalized to systems with even larger spin Chern numbers. For instance, one can consider two copies of the system concerned by us. By designing the coupling of the two layers to preserve the chiral symmetry, the number of MZMs per corner will be doubled.
In experiments, multiple MZMs will lead to a stronger zero-bias peak compared to a single MZM, and a lift of the chiral symmetry by applying a controllable external field can result in characteristic splitting features, which can be detected by scanning tunneling microscopy\cite{Liu2024MMZM}. Materials suitable for our proposal are abundant\cite{Bai2022DQSHE,Xue2023AQSHE,Cook2023HSCI,Liu2024SCI,Kang2024DQSHE}, like $\alpha$-Bi\cite{Bai2022DQSHE}, which
carry a spin Chern number $C_{\rm spin}=2$ and have a large bulk energy gap.
To conclude, our work reveals a route
to realize 2D SOTSCs with chiral-symmetry-protecting multifold MCMs.

\section*{Acknowledgments}
 The authors thank useful discussion with Shuxuan Wang and Junnan Lu. Zhiwei Yin and Shaolong Wan acknowledge the support by NSFC Grant No. 11275180. Haoshu Li acknowledges the support from the China Postdoctoral Science Foundation 2023M743398. Zhongbo Yan acknowledges the support from the National Natural Science Foundation of China (Grant No.
12174455) and the Natural Science Foundation of Guangdong Province (Grant No. 2021B1515020026).

\appendix
\section{Low-energy edge Hamiltonians for the first BdG Hamiltonian}
\label{sectin:appA}
Here we provide details of the derivation for the low-energy  Hamiltonians describing the other three edges.

To simplify the analysis, we consider the system occupies the whole upper half plane, so that the boundary corresponds to the edge II. Owing to the breaking of translation symmetry in the $y$ direction,  the Hamiltonian needs to be modified by replacing $k_y$ in Eq. (\ref{eq:decomposedH}) with the operator $-i \partial_y$. Accordingly, one has
\begin{equation}
\begin{aligned}
H_0(k_x,-i \partial_y)= & {\left[M\left(k_x\right)-t_y \partial_y^2\right] \tau_z \sigma_z-i v\left(k_x\right) \partial_y \tau_z \sigma_y s_y }\,, \\
H_1(k_x,- i \partial_y)= & -\mu \tau_z-\lambda\left(k_x^2+\partial_y^2\right) \tau_z \sigma_x \\
& +2 \eta k_x \tau_z \sigma_y s_x+\Delta_s \tau_y s_y \,,
\end{aligned}
\label{eq:ynormaledge}
\end{equation}
where $M\left(k_x\right)=m-2 t_x-2 t_y+t_x k_x^2$, $v\left(k_x\right)=2 \lambda k_x$. Under the boundary conditions $\psi_\alpha(0)=\psi_\alpha(+\infty)=0$,
one can also find that $H_0(k_x,-i \partial_y)$ supports four zero-energy solutions, whose wave functions are given by
\begin{equation}
\psi_\alpha(y)=\mathcal{N}_y \sin \left(\gamma_1 y\right) e^{-\gamma_2 y} e^{i k_x x} \zeta_\alpha\,,
\end{equation}
where the normalization constant
$\mathcal{N}_y=2\sqrt{|\gamma_2(\gamma_1^2+\gamma_2^2)/\gamma_1^2|}$, $ \gamma_1=\sqrt{-M(k_x)/t_y-v^2(k_x)/4t_y^2}$ and $\gamma_2=|{v(k_x)}/{2t_y}|$. The form of $\zeta_\alpha$ depends on the sign of $v(k_x)$ and is determined by the constraint $\sigma_x s_y\zeta_\alpha=\operatorname{sgn}(v(k_x))\zeta_\alpha$. When $k_x>0$, $\sigma_x s_y\zeta_\alpha=\zeta_\alpha$, eigenvectors are chosen as $\zeta_\alpha^+=\chi_\alpha^+$. When $k_x<0$, $\sigma_x s_y\zeta_\alpha=-\zeta_\alpha$, eigenvectors are chosen as $\zeta_\alpha^-=\chi_\alpha^-$. The matrix elements of the perturbative Hamiltonian $H_1(k_x,- i\partial_y)$ in the basis of these four zero-energy boundary states are given by
\begin{equation}
H_{\mathrm{II}, \alpha \beta}\left(k_x\right)=\int_0^{+\infty} d y \psi_\alpha^*(y) H_2\left(k_x,-i \partial_y\right) \psi_\beta(y) .
\end{equation}
After a straightforward calculation, we obtain the low-energy effective Hamiltonian for  edge II{\small
\begin{equation}
    H_{\rm II} (k_x)=\left\{
    \begin{aligned}
        &-\lambda\left(\frac{M(k_x)}{t_y}+k_x^2\right)\tau_z s_z + \Delta_s\tau_y s_z-2\eta k_x\tau_z s_y,\\
         & \quad \quad \quad \quad \quad \quad \quad \text{for } k_x > 0; \\
        &\lambda\left(\frac{M(k_x)}{t_y}+k_x^2\right)\tau_z s_z + \Delta_s\tau_y s_z+2\eta k_x\tau_z s_y,\\
         & \quad \quad \quad \quad \quad \quad \quad \text{for } k_x < 0. \\
    \end{aligned}
    \right .
\end{equation}}
Expanding around the two valleys at $K_{x,\pm}=\pm k_0$ ($K_{x,\pm}$ are solutions of the equation $M(k_x)/t_y+k_x^2=0$) to first order in momentum, we obtain the Hamiltonians for the two valleys
\begin{equation}
    \begin{aligned}
        H_{\rm  II}^+(q_x) = -\Lambda_{\rm  II} q_x\tau_z s_z + \Delta_s\tau_y s_z - 2\eta k_0\tau_z s_y,\\
        H_{\rm  II}^-(q_x) = -\Lambda_{\rm  II} q_x\tau_z s_z + \Delta_s\tau_y s_z - 2\eta k_0\tau_z s_y,
    \end{aligned}
\end{equation}
where $\Lambda_{\rm  II}=2\lambda(t_x+t_y)k_0\big/t_y$, and
$q_{x}$ denotes a small momentum measured from the concerned valley.

Similarly, to determine the low-energy Hamiltonian on edge III, we consider that the system occupies the whole left-half plane, so that the boundary at $x=0$ corresponds to edge III. For this case, the continuum Hamiltonian
 is identical to that on edge I, but the boundary conditions become $\psi_\alpha(0) = \psi_\alpha(-\infty) = 0$. It is straightforward to find that $H_0$ also supports four zero-energy eigenstates, whose wave functions  are similarly given by
\begin{equation}
\psi_\alpha(x)=\mathcal{N}_x \sin \left(\kappa_1 x\right) e^{\kappa_2 x} e^{i k_y y} \tilde{\chi}_\alpha \,.
\end{equation}
Here the expressions of $\mathcal{N}_x$, $\kappa_1$ and
$\kappa_2$ are the same as in the main text. Similarly, the spinor $\tilde{\chi}_\alpha$ depends on $k_{y}$. When $k_y>0$, $\sigma_x s_y\tilde\chi_\alpha=-\tilde\chi_\alpha$, accordingly the four eigenvectors $\tilde\chi_\alpha$ can be chosen as
\begin{equation}
\begin{aligned}
& \tilde\chi_1^+=\;\ket{\tau_z=+1}\otimes\ket{\sigma_x=-1}\otimes\ket{s_y=+1}\,, \\
& \tilde\chi_2^+=\;\ket{\tau_z=+1}\otimes\ket{\sigma_x=+1}\otimes\ket{s_y=-1}\,, \\
& \tilde\chi_3^+=\;\ket{\tau_z=-1}\otimes\ket{\sigma_x=-1}\otimes\ket{s_y=+1}\,, \\
& \tilde\chi_4^+=\;\ket{\tau_z=-1}\otimes\ket{\sigma_x=+1}\otimes\ket{s_y=-1}\,.
\end{aligned}
\end{equation}
When $k_y<0$, $\sigma_x s_y\tilde\chi_\alpha=\tilde\chi_\alpha$, hence the four eigenvectors $\tilde\chi_\alpha$ can be chosen as
\begin{equation}
\begin{aligned}
& \tilde\chi_1^-=\;\ket{\tau_z=+1}\otimes\ket{\sigma_x=+1}\otimes\ket{s_y=+1}\,, \\
& \tilde\chi_2^-=\;\ket{\tau_z=+1}\otimes\ket{\sigma_x=-1}\otimes\ket{s_y=-1}\,, \\
& \tilde\chi_3^-=\;\ket{\tau_z=-1}\otimes\ket{\sigma_x=+1}\otimes\ket{s_y=+1}\,, \\
& \tilde\chi_4^-=\;\ket{\tau_z=-1}\otimes\ket{\sigma_x=-1}\otimes\ket{s_y=-1}\,.
\end{aligned}
\end{equation}
By projecting $H_1(-i \partial_x, k_y)$ onto the subspace spanned by the four zero-energy states, we obtain the low-energy effective Hamiltonian
\begin{equation}
    H_{\rm III} (k_y)=\left\{\begin{aligned}
        -\lambda\left(\frac{M(k_y)}{t_x}+k_y^2\right)\tau_z s_z + \Delta_s\tau_y s_z,  \quad  &  k_y>0,\\
        \lambda\left(\frac{M(k_y)}{t_x}+k_y^2\right)\tau_z s_z + \Delta_s\tau_y s_z,  \quad & k_y<0.
    \end{aligned}
    \right .
\end{equation}
Expanding around the two valleys, one can find that the two valley Hamiltonians are of the form
\begin{equation}
\begin{aligned}
& H_{\mathrm{III}}^{+}\left(q_y\right)=-\Lambda_{\rm  III} q_y \tau_z s_z+\Delta_s \tau_y s_z \\
& H_{\mathrm{III}}^{-}\left(q_y\right)=-\Lambda_{\rm  III} q_y \tau_z s_z+\Delta_s \tau_y s_z,
\end{aligned}
\end{equation}
where $\Lambda_{\rm  III}=2\lambda(t_x+t_y)k_0\big/t_x=\Lambda_{\rm  I}$, and $q_{y}$ denotes a small momentum measured from the concerned valley.

Finally,  we consider that the system occupies the whole lower-half plane so that the boundary corresponds to edge IV. According to the boundary conditions $\psi_\alpha(0) = \psi_\alpha(-\infty) = 0$, one can also find that $H_0(k_x, -i\partial_y)$ in Eq. (\ref{eq:ynormaledge}) has four zero-energy solutions, whose wave functions are  given by
\begin{equation}
\psi_\alpha(y)=\mathcal{N}_y \sin \left(\gamma_1 y\right) e^{\gamma_2 y} e^{i k_x x} \tilde\zeta_\alpha \,.
\end{equation}
Similarly, when $k_x>0$, $\sigma_x s_y\tilde\zeta_\alpha=-\tilde\zeta_\alpha$, so one can choose $
\tilde\zeta_1^+=\tilde\chi_\alpha^+$. When $k_x<0$, $\sigma_x s_y\tilde\zeta_\alpha=\tilde\zeta_\alpha$, then one can choose $\tilde\zeta_1^-=\tilde\chi_\alpha^-$. By projecting $H_1$ onto the subspace spanned by these four zero-energy states, we obtain{\small
\begin{equation}
    H_{\rm IV} (k_x)=\left\{
    \begin{aligned}
        & \lambda\left(\frac{M(k_x)}{t_y}+k_x^2\right)\tau_z s_z + \Delta_s\tau_y s_z+2\eta k_x\tau_z s_y,\\
        & \quad \quad \quad \quad \quad \quad \quad \text{for } k_x > 0; \\
        &-\lambda\left(\frac{M(k_x)}{t_y}+k_x^2\right)\tau_z s_z + \Delta_s\tau_y s_z-2\eta k_x\tau_z s_y, \\
         & \quad \quad \quad \quad \quad \quad \quad \text{for } k_x < 0.\\
    \end{aligned}
    \right .
\end{equation}}
Again expanding around the two valleys, one obtains
\begin{equation}
    \begin{aligned}
        H_{\rm  IV}^+(q_x) = \Lambda_{\rm  IV} q_x\tau_z s_z + \Delta_s\tau_y s_z + 2\eta k_0\tau_z s_y\\
        H_{\rm  IV}^-(q_x) = \Lambda_{\rm  IV} q_x\tau_z s_z + \Delta_s\tau_y s_z + 2\eta k_0\tau_z s_y,
    \end{aligned}
\end{equation}
where $\Lambda_{\rm  IV}=2\lambda(t_x+t_y)k_0\big/t_y=\Lambda_{\rm  II}$.

\section{Low-energy edge Hamiltonians for the second BdG Hamiltonian}
\label{sectin:appB}

The process to derive the low-energy edge Hamiltonians for the second BdG Hamiltonian is completely the same.
The first step is also to divide the Hamiltonian in Eq.(\ref{eq:SPmodel}) into two parts, $H=H_{0}+H_{1}$, where
\begin{equation}
    \begin{aligned}
    H_0(k_x,k_y) = \; & (m-2 t_x \cos k_x-2 t_y \cos k_y) \tau_z \sigma_z \notag \\
    & + 2 \lambda \sin k_x \sin k_y \tau_z \sigma_y s_y \\
    H_1(k_x,k_y) = \; & 2 \lambda(\cos k_x-\cos k_y) \tau_z \sigma_x  + \Delta_s \tau_y  s_y\\
    &+ 2 \Delta_p \sin k_x \tau_y.
    \end{aligned}
\end{equation}
Here we also consider $\mu=0$ for simplicity. Without loss of generality, we assume that the band inversion occurs at $(k_x,k_y) = (0,0)$. By an expansion of these two Hamiltonians around this point to the second order of the momentum, one obtains their continuum counterparts, which read
\begin{equation}
    \begin{aligned}
    H_0(k_x,k_y) = \; & (m-2 t_x-2 t_y+t_x k_x^2+t_y k_y^2) \tau_z \sigma_z  \notag \\
    & + 2 \lambda k_x k_y \tau_z \sigma_y s_y \\
    H_1(k_x,k_y) = \; &- \lambda(k_x^2-k_y^2) \tau_z \sigma_x + \Delta_s \tau_ys_y+ 2 \Delta_p k_x \tau_y.
    \end{aligned}
\end{equation}

In Appendix \ref{sectin:appA}, we have already obtained the zero-energy solutions for $H_0(k_x,k_y)$ on each edge, and the difference in the two $H_1(k_x,k_y)$ lies only in the term $2 \Delta_p k_x \tau_y$.
Therefore, we only need to determine the result of this term when it is projected onto the subspace spanned by the zero-energy states.

For edge I and edge III, it is easy to check that the matrix elements generated by the $\Delta_{p}$ term are equal to zero. For edge II,
the matrix elements generated by the $\Delta_{p}$ term are given by
\begin{equation}
    H^p_{\mathrm{II}, \alpha \beta}\left(k_x\right) = \int_0^{+\infty} d y \psi_\alpha^*(y) \left(2\Delta_p k_x \tau_y \right) \psi_\beta(y) \,.
\end{equation}
After some simple calculations, we obtain
\begin{equation}
    H^p_{\mathrm{II}}\left(k_y\right) = 2\Delta_p k_x \tau_y \,.
\end{equation}
Thus, the low-energy effective Hamiltonian on edge II can be expressed as
\begin{equation}
    H_{\rm II} (k_x) = \left\{
    \begin{aligned}
        & -\lambda\left(\frac{M(k_x)}{t_y} + k_x^2\right) \tau_z s_z + \Delta_s \tau_y s_z + 2\Delta_p k_x \tau_y, \notag \\
        & \quad \quad \quad \quad \quad \quad \quad \text{for } k_x > 0; \\
        & \lambda\left(\frac{M(k_x)}{t_y} + k_x^2\right) \tau_z s_z + \Delta_s \tau_y s_z + 2\Delta_p k_x \tau_y, \notag \\
        & \quad \quad \quad \quad \quad \quad \quad \text{for } k_x < 0.
    \end{aligned}
    \right.
\end{equation}
where \(M(k_x) = m - 2t_x - 2t_y + t_x k_x^2\).

Similarly, for edge IV, the low-energy effective Hamiltonian is given by
\begin{equation}
    H_{\rm IV} (k_x) = \left\{
    \begin{aligned}
        & \lambda\left(\frac{M(k_x)}{t_y} + k_x^2\right) \tau_z s_z + \Delta_s \tau_y s_z + 2\Delta_p k_x \tau_y, \notag \\
        & \quad \quad \quad \quad \quad \quad \quad \text{for } k_x > 0; \\
        & -\lambda\left(\frac{M(k_x)}{t_y} + k_x^2\right) \tau_z s_z + \Delta_s \tau_y s_z + 2\Delta_p k_x \tau_y, \notag \\
        & \quad \quad \quad \quad \quad \quad \quad \text{for } k_x < 0.
    \end{aligned}
    \right.
\end{equation}

On each edge, there are two valleys in
the energy spectrum. Expanding the
Hamiltonian around the valleys, one
can obtain linear Dirac Hamiltonians
which are convenient for the analysis of the edge topology in terms of
the Jackiw-Rebbi theory. We label
the valley with positive (negative) momentum as $+$ $(-)$, then the low-energy Dirac Hamiltonian describing the four edges are given by
\begin{equation}
    \begin{aligned}
        H_{\rm I}^{\pm}(q_y) &= \Lambda_{\rm I} q_y \tau_z s_z + \Delta_s \tau_y s_z, \\
        H_{\rm II}^{\pm}(q_x) &= -\Lambda_{\rm II} q_x \tau_z s_z + \Delta_s \tau_y s_z \pm2\Delta_p k_0 \tau_y, \\
        H_{\mathrm{III}}^{\pm}(q_y) &= -\Lambda_{\rm III} q_y \tau_z s_z + \Delta_s \tau_y s_z, \\
        H_{\rm IV}^{\pm}(q_x) &= \Lambda_{\rm IV} q_x \tau_z s_z + \Delta_s \tau_y s_z \pm 2\Delta_p k_0 \tau_y,
    \end{aligned}
\end{equation}
where $\Lambda_{\rm I} = \Lambda_{\rm III} = 2\lambda(t_x + t_y)k_0 / t_x$, $\Lambda_{\rm II} = \Lambda_{\rm IV} = 2\lambda(t_x + t_y)k_0 / t_y$, and $k_0 = \sqrt{(2t_x + 2t_y - m) / (t_x + t_y)}$.

Introducing the boundary coordinate $l$, we then reach the compact form
\begin{equation}
        H_{\text{edge}}^{\pm} = i \Lambda(l) \tau_z s_z \partial_l + \Delta_s \tau_y s_z \pm M_p(l) \tau_y,
\end{equation}
Here, when $l$ is on the edge I (II, III, IV), $\Lambda(l)$ is equal to $\Lambda_{\rm I}$ ($\Lambda_{\rm II}, \Lambda_{\rm III}, \Lambda_{\rm IV}$), and $M_p(l)$ is equal to $0$ ($2\Delta_p k_0, 0, 2\Delta_p k_0)$.

\section{Calculation of the bulk topological invariant}\label{sectin:appC}

For two-dimensional higher-order topological phases with chiral symmetry, it has been shown that the number of zero-energy states per corner can be characterized by the multipole chiral numbers  $N_{xy}$ \cite{Benalcazar2022CSHOTI}. This invariant is defined based on the real-space Hamiltonian with periodic boundary conditions. To be specific, when a Hamiltonian has chiral symmetry, there exists a unitary operator $S$ anticommuting with $H(\bk)$, i.e.,
\begin{equation}
    \mathcal{S} H(\bk) \mathcal{S}^{-1} = -H(\bk).
\end{equation}

The presence of chiral symmetry implies that the Hilbert space of the Hamiltonian can be divided into two subspaces, denoted as sublattice $A$ and sublattice $B$. Accordingly, one can introduce two quadrupole momentums associated with them\cite{Benalcazar2017prb, Benalcazar2017},   denoted by $\Bar{Q}^A_{xy}$ and $\Bar{Q}^B_{xy}$. In the occupied-state basis, these two quadrupole moments can be expressed as
\begin{align}
    \Bar{Q}^{S}_{xy,mn} = 2 \sum_{\mathbf{R},\alpha \in S} \braket{\psi_m|\mathbf{R},\alpha} \Exp \left( -i \frac{2\pi xy}{L_xL_y}\right) \braket{\mathbf{R},\alpha| \psi_n} \, ,
    \label{eq: Qocc}
\end{align}
where $S=\{A,B\}$, $L_{x}$ and $L_{y}$ denote the number of unit cells in $x$ and $y$ directions, $\ket{\mathbf{R},\alpha} = c_{\mathbf{R},\alpha}^\dagger \ket{0}$ with $c_{\mathbf{R},\alpha}^\dagger$ being the fermionic creation operator for the $\alpha$ orbital in the unit cell at $\mathbf{R} = (x, y)$, and     $|\psi_m\rangle$ represents the $m$-th occupied state of the Hamiltonian.  In terms of the two quadrupole moments $\Bar{Q}^A_{xy}$ and $\Bar{Q}^B_{xy}$, two integer-valued quantities $N^A_{xy}$ and $N^B_{xy}$ can be defined as
\begin{align}
    N^{S}_{xy} = \frac{1}{2\pi i} \Tr \log\left(\Bar{Q}_{xy}^{S}\right) , \quad S=A,B.
    \label{eq:invAB}
\end{align}
In chiral symmetric systems, $N^A_{xy}$ and $N^B_{xy}$ themselves are not topological invariants, but their difference
\begin{align}
    N_{xy} = N^A_{xy} - N^B_{xy} = \frac{1}{2\pi i} \Tr \log\left(\Bar{Q}_{xy}^A \Bar{Q}_{xy}^{B\dagger}\right)
    \label{eq: topo_inv_N}
\end{align}
is a topological invariant. In Ref. \cite{Benalcazar2022CSHOTI}, another form of $\Bar{Q}^S_{xy}$ is given, which is more suitable for numerical calculations. In the following, we describe this equivalent form of $\Bar{Q}^S_{xy}$.

For a chiral symmetric system with balanced sublattices, the chiral symmetry operator can also be made to take the
specific form $\mathcal{S} = \tau_z$ by
appropriately choosing the basis. Interestingly, when  $\mathcal{S} = \tau_z$, the Hamiltonian will take a block-off-diagonal form
\begin{equation}
    H = \begin{pmatrix}
        0 & h \\
        h^\dagger & 0
    \end{pmatrix}.
\end{equation}
According to the two eigenvalues of $\mathcal{S} $,  the lattice can be divided into two sublattices, $A$  and $B$ . In this basis, the eigenstates of $H$ can be written as $\ket{\psi_n} = \left(\frac{1}{\sqrt{2}}\right) \left( \psi_n^A, \psi_n^B \right)^T$, where $\psi_n^A, \psi_n^B$ represent the normalized vectors in the $A$ and $B$ subspaces, respectively. In fact, they are the eigenstates of the Hermitian operators $h h^\dagger$ and $h^\dagger h$, respectively. The vectors from different subspaces can be related through the singular value decomposition of $h$
\begin{equation}
    h = U_A \Sigma U_B^\dagger \,,
    \label{eq: SVDh}
\end{equation}
where $U_S$ $(S = A, B)$ are the unitary matrices formed by $\{ \psi_n^S \}$, i.e., $U_S = \left( \psi_1^S, \psi_2^S, \cdots, \psi_{N_S}^S \right)$, and $\Sigma$ is the diagonal matrix composed of the singular values of $h$.

The quadrupole operators defined on each sublattice without projecting onto the occupied states subspace (note that $\Bar{Q}_{xy}^{S}$ defined in Eq.~(\ref{eq: Qocc}) are projected operator acting on the occupied states subspace) are
\begin{equation}
    Q_{xy}^S = \sum_{\mathbf{R},\alpha \in S} \ket{\mathbf{R},\alpha} \Exp \left( -i\frac{2\pi xy}{L_xL_y}\right) \bra{\mathbf{R},\alpha} \,.
    \label{eq: Qform}
\end{equation}
The projected operators $\Bar{Q}_{xy}^S$ can be expressed in terms of the unprojected operator $Q_{xy}^S$  and the transformation matrices of the singular value decomposition in Eq.~(\ref{eq: SVDh}), i.e.,
\begin{align}
    \Bar{Q}_{xy}^S = U_S^\dagger Q_{xy}^S U_S, (S = A, B) \, .
    \label{eq: proQform}
\end{align}
By using Eq.~(\ref{eq: topo_inv_N}, \ref{eq: SVDh}, \ref{eq: Qform} and \ref{eq: proQform}), the value of the topological invariant $N_{xy}$ can be calculated numerically.

\bibliography{dirac}

\end{document}